\documentclass[
    aip,
    jcp,
    reprint,
    amsmath,
    amssymb,
    floatfix
]{revtex4-2}

\usepackage{graphicx}
\usepackage{dcolumn}
\usepackage{bm}
\usepackage{chemformula}
\usepackage{siunitx}
\usepackage{youngtab}
\usepackage{tabularx}
\usepackage{array}
\usepackage{needspace}
\usepackage{xcolor}
\usepackage[normalem]{ulem}

\allowdisplaybreaks

\begin{document}

\title{Origin of Long-Lived Nuclear Spin States and Coherences in Aliphatic Chains Revealed by Relaxation Theory}

\author{Danil A. Markelov}
\email{danil.markelov@ens.psl.eu}
\affiliation{Chimie Physique et Chimie du Vivant (CPCV, UMR 8228), D\'epartement de Chimie, \'{E}cole Normale Sup\'erieure, PSL University, Sorbonne Universit\'e, Paris 75005, France}

\author{Kirill F. Sheberstov}
\email{kirill.sheberstov@ens.psl.eu}
\affiliation{Chimie Physique et Chimie du Vivant (CPCV, UMR 8228), D\'epartement de Chimie, \'{E}cole Normale Sup\'erieure, PSL University, Sorbonne Universit\'e, Paris 75005, France}

\date{\today}

\begin{abstract}
Delocalized long-lived states (LLSs) and collective zero-quantum long-lived coherences
(LLCs) in aliphatic chains provide a promising route for preserving
nuclear spin order with lifetimes that exceed the conventional
$T_1$ and $T_2$ relaxation times, respectively. Their extended lifetimes make them attractive for applications including hyperpolarization storage, ligand-observed drug screening based on the loss of longevity upon binding to a target protein, and quantum information processing exploiting the collective properties of many-body spin systems. Although LLSs and LLCs have
been observed experimentally in methylene networks, their
origin and general structure in chains of arbitrary length have remained unclear. Here we show
that these relaxation-protected modes follow directly from Redfield
relaxation theory. Specifically, we construct the long-lived subspace, i.e., the zero-eigenvalue
subspace of the relaxation superoperator associated with
the dominant intra-pair dipole--dipole relaxation mechanism. The long-lived subspace contains $2^N-1$ independent non-trivial operators, which excludes the identity operator, where $N>1$ is the number of $-\mathrm{CH}_2-$ groups in the chain. In achiral molecules, conservation of the global intra-pair permutation parity restricts experimental access to at most $2^N-2$ of these operators, whereas in chiral molecules this parity is not conserved, making up to $2^N-1$ long-lived operators accessible. We further develop a general framework for constructing both LLSs and LLCs in aliphatic chains containing an arbitrary number of $-\mathrm{CH}_2-$ groups in achiral molecules, and illustrate the approach explicitly for chains with $N=2$, 3, and 4 methylene groups.
\end{abstract}

\maketitle

\section{Introduction}

In nuclear magnetic resonance (NMR), the lifetime of nuclear
magnetization is usually limited by the longitudinal
relaxation time, $T_1$. Long-lived states (LLSs) form an
important exception: their lifetimes, $T_{\mathrm{LLS}}$,
can exceed $T_1$ by orders of magnitude and thereby
preserve non-equilibrium magnetization over extended
periods of time. \cite{carravetta2004long,meier2013long,pileio2020long,vasos2009long,stevanato2015nuclear,sheberstov2019generating,sheberstov2018cis,pileio2008long} This property makes LLSs particularly
attractive for storing hyperpolarization, \cite{pileio2010storage,kiryutin2024photo} probing slow
molecular processes, \cite{ahuja2009diffusion,pileio2015real,sarkar2008measurement,cavadini2005slow,sarkar2007singlet} and detecting weak molecular
interactions. \cite{kozinenko2024exploring,butikofer2026combining,salvi2012boosting,buratto2014drug,geniman2024consistent} Closely related
zero-quantum long-lived coherences (LLCs) can also
survive longer than the transverse relaxation time,
$T_2$.\cite{bornet2011ultra,sarkar2010long,pileio2020long} Moreover, as zero-quantum coherences, they are insensitive to magnetic-field inhomogeneity and can therefore give rise to exceptionally narrow NMR lines.  \cite{sarkar2010long,sadet2026long,sarkar2011long,chinthalapalli2012ultrahigh,sheberstov2019excitation}

The simplest and most well-known example of a
long-lived state is the singlet order in a pair of coupled
spin-$1/2$ nuclei. \cite{pileio2010relaxation,levitt2012singlet,tayler2011singlet,pileio2007j} For two spins, the singlet--triplet basis
can be written in terms of the Zeeman product states as
\begin{equation}
\begin{aligned}
|T_{+1}\rangle &= |\alpha\alpha\rangle,
&
|T_0\rangle &= \frac{|\alpha\beta\rangle+|\beta\alpha\rangle}{\sqrt{2}},
\\
|T_{-1}\rangle &= |\beta\beta\rangle,
&
|S_0\rangle &= \frac{|\alpha\beta\rangle-|\beta\alpha\rangle}{\sqrt{2}}.
\label{eq:singlet}
\end{aligned}
\end{equation}
The singlet order is represented by the
population imbalance,
$
\hat{\text{SO}}
=
|S_0\rangle\langle S_0|
-
\frac{1}{3}
\sum_{i=-1}^{1}
|T_i\rangle\langle T_i|,$
where $|S_0\rangle$ is the antisymmetric singlet state
and $|T_i\rangle$ are the three symmetric triplet states.
The long lifetime of this operator follows from symmetry,
since transitions between different symmetry manifolds
are forbidden. \cite{freeman1970high} From a similar symmetry considerations it turns out that the fluctuating intra-pair dipole--dipole interaction cannot convert singlet population into triplet population, so that $\hat{\text{SO}}$ belongs to the zero-eigenvalue subspace of the corresponding relaxation superoperator. For this reason, many experimental approaches have been developed for preparing, sustaining, and detecting long-lived singlet order,\cite{carravetta2005theory,pileio2009theory,gopalakrishnan2006lifetimes,kiryutin2016nuclear,vinogradov2008hyperpolarized,elliott2019field} including spin-lock-induced
crossing (SLIC),\cite{devience2013preparation,sabba2026error} adiabatic-passage spin order conversion
(APSOC),\cite{rodin2018using,kiryutin2013manipulating,kiryutin2015long,pravdivtsev2016robust} magnetization-to-singlet conversion (M2S), \cite{pileio2010storage,bengs2020generalised} and
other NMR pulse sequences. \cite{hullamballi2025quantum,heramun2026spinor,sabba2022symmetry,bengs2023aharonov}

Long-lived spin order is not restricted to isolated
spin pairs. \cite{vinogradov2007long,hogben2011multiple} It can also arise in symmetric multi-spin systems,
where spin states belonging to different irreducible
representations of the molecular symmetry group may be weakly
connected by the dominant relaxation mechanisms. \cite{meier2013long,zhukov2019assessment,dumez2015theory,feng2012accessing, franzoni2012long} Another important class of multi-spin systems is provided by aliphatic
chains. It has been shown that chains containing several
methylene groups, $(-\mathrm{CH}_2-)_N$, can support delocalized long-lived modes
involving nuclear spins throughout the entire aliphatic chain. \cite{sonnefeld2022long,sonnefeld2022polychromatic,razanahoera2023paramagnetic,razanahoera2024hyperpolarization,wiame2025long,vandyck2026excitation,butikofer2026combining,xuan2026effects} In aliphatic chains, each methylene group contains a pair of
geminal protons, making the intra-pair dipole--dipole interaction
within each $-\mathrm{CH}_2-$ group the dominant relaxation
channel, while weaker inter-pair dipole--dipole interactions provide
slower relaxation pathways between different groups. Experiments on
aliphatic chains have demonstrated polychromatic SLIC-based excitation of not only
delocalized long-lived states but also
collective long-lived coherences. \cite{sheberstov2024collective,sheberstov2025aliphatic} These results show that aliphatic chains, which are common
structural motifs in many molecules, including biologically
relevant compounds, provide a natural platform for preparing
long-lived spin order.

However, the origin of these delocalized LLSs and collective LLCs in
multi-spin chains is less evident than in the two-spin
case. For a single spin pair, the structure of the
long-lived operator follows directly from the permutation
symmetry of the nuclear spins, because the dipole--dipole
relaxation superoperator cannot induce transitions
between different symmetry manifolds. For a chain of
several $-\mathrm{CH}_2-$ groups, the situation is more
complex. In particular, LLSs and LLCs in aliphatic chains have been observed experimentally, but their status as relaxation-protected modes has not been derived from the first principles.

In this work, we derive the structure of long-lived operators in aliphatic chains directly from Redfield relaxation theory. Starting from the intra-pair dipole--dipole relaxation superoperator, we derive the structure of the zero-eigenvalue subspace that contains the long-lived operators. This leads to a general procedure for constructing the multi-spin LLSs and LLCs basis for a molecule containing an arbitrary number of $-\mathrm{CH}_2-$ groups. We formally demonstrate that an aliphatic chain $(-\mathrm{CH}_2-)_N$ supports $2^N-1$ independent non-trivial long-lived operators, excluding the identity operator. Our analysis focuses primarily on achiral molecules, for which conservation of the global intra-pair permutation parity restricts the number of experimentally excitable operators to at most $2^N-2$. We briefly discuss how this restriction is lifted in chiral molecules, where up to $2^N-1$ non-trivial operators can, in principle, become experimentally accessible.
We also show that long-lived coherences in aliphatic chains arise from the same
relaxation-protected subspace as the long-lived states. Thus, LLCs are not additional phenomenological
objects, but follow naturally from the Redfield description. We illustrate the construction explicitly for chains with $N=2$, $N=3$, and $N=4$ methylene groups.

\section{Theory}

\subsection{The Dipole-Dipole Relaxation Superoperator}
Let us start by considering the relaxation superoperator. We assume that the aliphatic spin chain contains $N$ $-\mathrm{CH}_2-$ groups. For a given spin pair $(i,j)$ of the chain, the corresponding dipole-dipolar relaxation superoperator in the Redfield's framework can be written as \cite{pileio2010relaxation}
\begin{equation}
\hat{\hat{\mathbf{R}}}^{(i,j)}
=
-a_{\mathrm{DD}}^{(i,j)}
\sum_{m=-2}^{2}
(-1)^m
\hat{\hat{\mathbf{T}}}_{2,-m}^{(i,j)}
\hat{\hat{\mathbf{T}}}_{2,m}^{(i,j)} ,
\label{eq:Rij-DD}
\end{equation}
where
\begin{equation}
a_{\mathrm{DD}}^{(i,j)}
=
\frac{6\tau_c b^{(i,j)}}{5},
\qquad
b^{(i,j)}
=
\left(
\frac{\mu_0}{4\pi}
\right)^2
\frac{[\gamma^{(i)}]^2[\gamma^{(j)}]^2\hbar^2}
{\left[r^{(i,j)}\right]^6}.
\label{eq:aDD}
\end{equation}
Here, $\tau_c$ is the rotational correlation time; $r^{(i,j)}$ is the distance between spins $i$ and $j$; $\gamma^{(i)}$, $\gamma^{(j)}$ are the gyromagnetic ratios; and $\hat{\hat{\mathbf{T}}}_{2,m}^{(i,j)}$ denotes the commutation superoperator associated with the rank-two irreducible spherical superoperators of the dipole--dipole interaction for this spin pair. Each superoperator acts as a commutator, namely
\begin{equation}
\hat{\hat{\mathbf{T}}}_{2,m}^{(i,j)}\hat{L}
=
\left[
\hat{T}_{2,m}^{(i,j)},\hat{L}
\right],
\label{eq:T-superoperator-commutator}
\end{equation}
where $\hat{L}$ is an arbitrary spin operator, while $\hat{T}_{2,m}^{(i,j)}$ with a single hat denotes the rank-two irreducible spherical tensor operator for the spin pair $(i,j)$. Its explicit form is given below.

In what follows, we neglect cross-correlations between different dipole--dipole interaction sites. This approximation is justified by the fact that the corresponding cross-correlation spectral densities contain the angular factor
$\
\frac{1}{2}
\left[
3\cos^2\theta^{(i,j;k,l)}-1
\right],
$
where $\theta^{(i,j;k,l)}$ is the angle between the two dipolar vectors $\mathbf{r}^{(i,j)}$ and $\mathbf{r}^{(k,l)}$. \cite{kumar2000cross} For rapidly rotating $-\mathrm{CH}_2-$ groups, the relative orientation of different dipolar vectors is efficiently averaged, which suppresses these cross-correlation terms. Therefore, the dominant contribution to the relaxation superoperator is expected to arise from the auto-correlation terms of dipole--dipole interactions.

Neglecting the cross-correlation terms, the relaxation superoperator is given by the sum over all spin pairs,
\begin{equation}
\hat{\hat{\mathbf{R}}}
=
\sum_{i<j}
\hat{\hat{\mathbf{R}}}^{(i,j)}.
\label{eq:R-total}
\end{equation}
It is convenient to split this superoperator into intra-pair and inter-pair contributions,
\begin{equation}
\hat{\hat{\mathbf{R}}}
=
\hat{\hat{\mathbf{R}}}_{\mathrm{intra}}
+
\hat{\hat{\mathbf{R}}}_{\mathrm{inter}}.
\label{eq:R-intra-inter}
\end{equation}
In this work, we focus on the dominant intra-pair dipole-dipole relaxation, which is expected to be the main relaxation mechanism in aliphatic chains. For a chain of $(-\mathrm{CH}_2-)_N$ groups, $\hat{\hat{\mathbf{R}}}_{\text{intra}}$ is given by
\begin{equation}
\hat{\hat{\mathbf{R}}}_{\mathrm{intra}}
=
-
\sum_{k=1}^{N}a^{(2k-1,2k)}_{\mathrm{DD}}
\sum_{m=-2}^{2}
(-1)^m
\hat{\hat{\mathbf{T}}}_{2,-m}^{(2k-1,2k)}
\hat{\hat{\mathbf{T}}}_{2,m}^{(2k-1,2k)},
\label{eq:R-intra-compact}
\end{equation}
where the index $k=1 \ldots N$ numbers different $\text{CH}_2$ groups.

An operator $\hat{L}$ satisfying
\begin{equation}
\hat{\hat{\mathbf{R}}}_{\mathrm{intra}}\hat{L}=0
\label{eq:Rintra-kernel}
\end{equation}
represents a long-lived operator, because it is protected from the dominant intra-pair dipole--dipole relaxation mechanism. As a result, its lifetime is prolonged and is governed only by the weaker inter-pair relaxation contributions.

To analyze Eq.~(\ref{eq:Rintra-kernel}), we use the Hermitian conjugation property of irreducible spherical tensor superoperators, \cite{pyper1971theory,messiah1962quantum}
$
\left(
\hat{\hat{\mathbf{T}}}_{2,m}^{(2k-1,2k)}
\right)^{\dagger}
=
(-1)^m
\hat{\hat{\mathbf{T}}}_{2,-m}^{(2k-1,2k)}.
$
Substituting this relation into Eq.~\eqref{eq:R-intra-compact}, we obtain
\begin{equation}
\hat{\hat{\mathbf{R}}}_{\mathrm{intra}}
=
-
\sum_{k=1}^{N}a^{(2k-1,2k)}_{\mathrm{DD}}
\sum_{m=-2}^{2}
\left(
\hat{\hat{\mathbf{T}}}_{2,m}^{(2k-1,2k)}
\right)^{\dagger}
\hat{\hat{\mathbf{T}}}_{2,m}^{(2k-1,2k)} .
\label{eq:R-intra-positive-form}
\end{equation}

Here and below, we use the Hilbert--Schmidt scalar product,
\begin{equation}
\left(
\hat{A}
\middle|
\hat{B}
\right)
=
\operatorname{Tr}
\left\{
\hat{A}^{\dagger}\hat{B}
\right\},
\label{eq:HSproduct}
\end{equation}

which defines the norm of an operator
\begin{equation}
\|\hat{A}\|^2=
\left(
\hat{A}
\middle|
\hat{A}
\right).
\end{equation}

The key observation follows from evaluating the scalar product
$\left(\hat{L}\middle|\hat{\hat{\mathbf{R}}}_{\mathrm{intra}}\hat{L}\right)$ that
quantifies the relaxation-induced leakage from the operator $\hat{L}$ under the action of the intra--pair dipole--dipole relaxation superoperator:
\begin{align}
&\left(
\hat{L}
\middle|
\hat{\hat{\mathbf{R}}}_{\mathrm{intra}}\hat{L}
\right)
\nonumber \\
&= 
-
\sum_{k=1}^{N}a^{(2k-1,2k)}_{\mathrm{DD}}
\sum_{m=-2}^{2}
\left(
\hat{L}
\middle|
\left(
\hat{\hat{\mathbf{T}}}_{2,m}^{(2k-1,2k)}
\right)^{\dagger}
\hat{\hat{\mathbf{T}}}_{2,m}^{(2k-1,2k)}
\hat{L}
\right)
\nonumber\\
&=-
\sum_{k=1}^{N}a^{(2k-1,2k)}_{\mathrm{DD}}
\sum_{m=-2}^{2}
\left(
\hat{\hat{\mathbf{T}}}_{2,m}^{(2k-1,2k)}\hat{L}
\middle|
\hat{\hat{\mathbf{T}}}_{2,m}^{(2k-1,2k)}\hat{L}
\right)
\nonumber\\
&=-
\sum_{k=1}^{N}a^{(2k-1,2k)}_{\mathrm{DD}}
\sum_{m=-2}^{2}
\left\|
\hat{\hat{\mathbf{T}}}_{2,m}^{(2k-1,2k)}\hat{L}
\right\|^2.
\label{eq:Rintra-negative-semidefinite}
\end{align}
Therefore, if $\hat{L}$ is an eigenoperator of $\hat{\hat{\mathbf{R}}}_{\mathrm{intra}}$ corresponding to zero eigenvalue, then
\begin{equation}
\left(
\hat{L}
\middle|
\hat{\hat{\mathbf{R}}}_{\mathrm{intra}}\hat{L}
\right)
=0,
\end{equation}
which implies
\begin{equation}
\sum_{k=1}^{N}a^{(2k-1,2k)}_{\mathrm{DD}}
\sum_{m=-2}^{2}
\left\|
\hat{\hat{\mathbf{T}}}_{2,m}^{(2k-1,2k)}\hat{L}
\right\|^2
=0.
\label{eq:sum-norms-zero}
\end{equation}
Since all terms in this sum are non-negative and $a_{\text{DD}}^{(2k-1,2k)}>0$ for any $k$, the equality can hold if and only if each term vanishes separately:
\begin{equation}
\hat{\hat{\mathbf{T}}}_{2,m}^{(2k-1,2k)}\hat{L}=0
\text{ for all } m=-2,\ldots,2
\text{ and } k=1,\ldots,N.
\label{eq:annihilation-condition}
\end{equation}
Thus, an operator is protected from the dominant intra-pair dipole--dipole relaxation mechanism if and only if it is annihilated by every intra-pair rank-two dipole--dipole irreducible spherical tensor superoperator. This condition constitutes the fundamental origin of the long-lived operators (LLOs) constructed below. 

\subsection{The Local Long-Lived Operator}

The condition derived above provides a direct way to construct operators protected from the dominant intra-pair dipole--dipole relaxation mechanism. Let us fix a given $-\mathrm{CH}_2-$ group, indexed by $k$, and consider an operator $\hat{L}^{(2k-1,2k)}$ acting in the Hilbert space of the spin pair $(2k-1,2k)$. According to Eq.~\eqref{eq:annihilation-condition}, this operator must satisfy
\begin{equation}
\hat{\hat{\mathbf{T}}}_{2,m}^{(2k-1,2k)}
\hat{L}^{(2k-1,2k)}
=
0,
\qquad
m=-2,-1,0,1,2.
\label{eq:local-annihilation-condition}
\end{equation}
Equivalently, since the superoperators $\hat{\hat{\mathbf{T}}}_{2,m}^{(2k-1,2k)}$ act as commutators, one has
\begin{equation}
\left[
\hat{T}_{2,m}^{(2k-1,2k)},
\hat{L}^{(2k-1,2k)}
\right]
=
0,
\qquad
m=-2,-1,0,1,2.
\label{eq:local-commutator-condition}
\end{equation}

For compactness, we temporarily omit the pair index $(2k-1,2k)$ and denote the two spins in the pair by $\hat{\mathbf{I}}$ and $\hat{\mathbf{S}}$. The five rank-two tensor components of the dipole--dipole interaction then lead to the following set of commutation relations:
\begin{align}
\left[
\hat{I}_{+}\hat{S}_{+},
\hat{L}
\right]
&=0,
\nonumber\\
\left[
\hat{I}_{+}\hat{S}_{z}
+
\hat{I}_{z}\hat{S}_{+},
\hat{L}
\right]
&=0,
\nonumber\\
\left[
3\hat{I}_{z}\hat{S}_{z}
-
\hat{\mathbf{I}}\cdot\hat{\mathbf{S}},
\hat{L}
\right]
&=0,
\nonumber\\
\left[
\hat{I}_{-}\hat{S}_{z}
+
\hat{I}_{z}\hat{S}_{-},
\hat{L}
\right]
&=0,
\nonumber\\
\left[
\hat{I}_{-}\hat{S}_{-},
\hat{L}
\right]
&=0.
\label{eq:five-commutators-DD}
\end{align}
The five commutation relations in Eq.~(\ref{eq:five-commutators-DD}) correspond to the five irreducible spherical tensor components of the intra-pair dipole--dipole interaction, which form the rank-two dipolar interaction Hamiltonian and are conventionally identified with the (A)--(E) terms of Abragam's alphabet in magnetic resonance theory.\cite{abragam1961principles} Therefore, the desired local long-lived operator must commute with all five irreducible rank-two components of the intra-pair dipole--dipole interaction.

For a given pair of methylene protons, two linearly independent operators satisfying the five commutation relations in Eq.~\eqref{eq:five-commutators-DD} can be identified immediately. The first is the identity operator, $\hat{\mathbf{1}}$, which trivially commutes with all operators. The second is the singlet-state population operator,
$
|S_0\rangle
\langle S_0|.
$
The well-known rotational invariance  of the singlet state,\cite{pileio2010relaxation} $|S_0\rangle$, ensures that the corresponding singlet population operator $
|S_0\rangle
\langle S_0|.
$ commutes with all five rank-2 tensor components appearing in Eq.~\eqref{eq:five-commutators-DD}.

The identity operator can be decomposed into the singlet and triplet population projectors as
\begin{equation}
\hat{\mathbf{1}}
=
|S_0\rangle\langle S_0|
+
|T_{+1}\rangle\langle T_{+1}|
+
|T_0\rangle\langle T_0|
+
|T_{-1}\rangle\langle T_{-1}|.
\end{equation}
Consequently, the identity operator is not orthogonal to the singlet population operator
$|S_0\rangle\langle S_0|$. To obtain an orthogonal basis for the local zero-eigenvalue subspace, we therefore replace the identity operator by the projector onto the triplet manifold,
$
|T_{+1}\rangle\langle T_{+1}|
+
|T_0\rangle\langle T_0|
+
|T_{-1}\rangle\langle T_{-1}|.
$
The same local subspace is thus spanned by the mutually orthogonal operators,
$|S_0\rangle\langle S_0|$ and $
|T_{+1}\rangle\langle T_{+1}|
+
|T_0\rangle\langle T_0|
+
|T_{-1}\rangle\langle T_{-1}|
$.

It can be formally demonstrated that no other long-lived operators exist for a pair of spin-$1/2$ nuclei. To this end, an arbitrary operator is expanded in the singlet--triplet product-operator basis,
$
\left\{
|i\rangle\langle j|
\right\}, \text{where }
i,j\in
\left\{
S_0,T_{+1},T_0,T_{-1}
\right\},
$
which contains 16 basis operators. Substitution of this expansion into the five commutation relations in Eq.~\eqref{eq:five-commutators-DD} yields a linear system for the 16 expansion coefficients. Solving this system shows that the zero-eigenvalue subspace is indeed two-dimensional and is spanned by the singlet population operator and the sum of triplet population operators. Hence, no additional independent operators satisfy all five commutation relations.

Therefore, for each spin pair, the operator corresponding to the zero eigenvalue of the intra-pair dipole--dipole relaxation superoperator is spanned by the singlet population operator and the equally weighted triplet population operator. In other words, the local LLO is given by:
\begin{equation}
\hat{P}\!\left(S^{(2k-1,2k)}_0\right)
=
\left|
S^{(2k-1,2k)}_0
\right\rangle
\left\langle
S^{(2k-1,2k)}_0
\right|,
\end{equation}
and
\begin{equation}
\hat{P}\!\left(T^{(2k-1,2k)}\right)
=
\frac{1}{3}
\sum_{i=-1}^{1}
\left|
T_{i}^{(2k-1,2k)}
\right\rangle
\left\langle
T_{i}^{(2k-1,2k)}
\right|,
\label{eq:local-triplet-population}
\end{equation}
where $1/3$ in Eq.~\eqref{eq:local-triplet-population} is a normalization factor, as explained below. This result gives the elementary building block for the construction of the multiple LLS and LLC operators. 

It is important to note that the same local zero-eigenvalue subspace can also be represented using the intra-pair scalar spin product:

\begin{equation}
\begin{aligned}
\hat P(S_0^{(2k-1,2k)})
&=
\frac{1}{4}\hat{\mathbf 1}
-
\hat{\mathbf I}^{(2k-1)}\cdot \hat{\mathbf I}^{(2k)},
\\
\hat P(T^{(2k-1,2k)})
&=
\frac{1}{4}\hat{\mathbf 1}
+
\frac{1}{3}\hat{\mathbf I}^{(2k-1)}\cdot \hat{\mathbf I}^{(2k)}.
\end{aligned}
\end{equation}

Thus, together with the identity operator, the intra-pair scalar products of spin operators span the same two-dimensional local subspace as the singlet and averaged triplet population projectors. In previous works, LLSs and LLCs were often constructed numerically using such scalar-product operators. \cite{sonnefeld2022polychromatic,razanahoera2023paramagnetic}

In the following, we use the singlet--triplet population basis as the primary representation because it directly reflects the intra-pair permutation parity of the states involved and allows separation between the long-lived states and the long-lived coherences. Nevertheless, for comparison, the resulting LLO operators will also be expanded in the product-operator basis generated by the intra-pair scalar spin products.

\subsection{Combinatorial Structure of the Long-Lived Operators}

\begin{figure*}[t]
\centering
\includegraphics[width=12.5 cm]{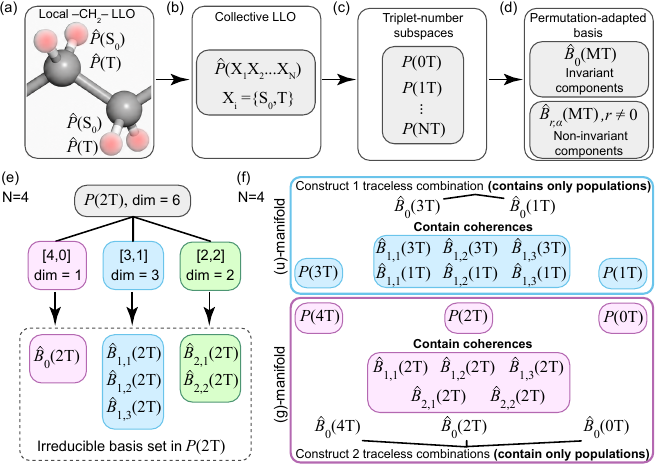}
\caption{
Schematic representation of the construction of long-lived operators (LLOs) in aliphatic spin chains. (a) The local LLO subspace of a single $-\mathrm{CH}_2-$ group is spanned by two basis operators, $\hat{P}(S_0)$ and $\hat{P}(T)$. (b) An operator corresponding to an LLO in an aliphatic spin chain containing $N$ $-\mathrm{CH}_2-$ groups is constructed as a product of either $\hat{P}(S_0)$ or $\hat{P}(T)$ within each group. (c) These products are classified according to the number of triplet populations $T$ they contain, thereby defining the triplet-number subspaces $P(MT)$. (d) Within each triplet-number manifold $P(MT)$, the operator basis is further decomposed into permutation-adapted components described by Young diagrams. (e) Example of the construction of permutation-adapted basis operators for $N=4$ within the $P(2T)$ subspace. Each Young diagram is labeled by the partition $[N-r,r]$. (f) Construction of an orthonormal LLO basis from irreducible basis operators adapted to the spin permutation symmetry. For the $N=4$ methylene chain shown here, the subspaces $P(3T)$ and $P(1T)$ belong to the ungerade ($u$) manifold, whereas $P(4T)$, $P(2T)$, and $P(0T)$ belong to the gerade ($g$) manifold. The operators $\hat{B}_{r,\alpha}$ with $r\neq0$ already represent LLOs containing both populations and coherences, i.e. LLSs and LLCs. The fully symmetric operators $\hat{B}_0(MT)$ are combined into traceless operators within each global intra-pair permutation parity manifold, $g$ or $u$. These combinations of $\hat{B}_0$ do not contain coherences and represent purely LLSs. 
}
\label{fig:1}
\end{figure*}

The local LLO obtained above shows that each $-\mathrm{CH}_2-$ group contributes two elementary zero-eigenvalue population operators: the singlet population $\hat{P}(S_0)$ and the equally weighted triplet population $\hat{P}(T)$, as shown in Fig.~\ref{fig:1}a. Therefore, for a system of coupled $(-\mathrm{CH}_2)_N-$ groups, the zero-eigenvalue subspace of the intra-pair relaxation superoperator is generated by product operators of the form (Fig.~\ref{fig:1}b)
\begin{equation}
\begin{aligned}
&\hat{\text{LLO}}
=
\prod_{k=1}^{N}
\hat{P}\!\left(X^{(2k-1,2k)}\right),\\
&
X^{(2k-1,2k)}\in\left\{S^{(2k-1,2k)}_0,T^{(2k-1,2k)}\right\}.
\end{aligned}
\label{eq:many-pair-product-operator}
\end{equation}
By construction, such operators satisfy Eq.~(\ref{eq:annihilation-condition}) and indeed belong to the zero-eigenvalue subspace of the intra-pair relaxation superoperator. Here and below, the symbol $T$ denotes the equally weighted triplet population at the corresponding methylene group. Additionally, in what follows, the superscripts denoting different $-\text{CH}_2-$ groups are omitted for compactness.

Thus, if a product of the operators in Eq.~(\ref{eq:many-pair-product-operator}) contains $M$ triplet labels, it implicitly contains the average over all magnetic sublevels of these triplets,
\begin{equation}
\begin{aligned}
&\hat{P}\!\left(
\underbrace{T\ldots T}_{M}
\underbrace{S_0\ldots S_0}_{N-M}
\right)
=
\\[-0.2em]
&\quad
\frac{1}{3^M}
\sum_{i_1,\ldots,i_M=-1}^{1}
\hat{P}\!\left(
\underbrace{T_{i_1}T_{i_2}\ldots T_{i_M}}_{M}
\underbrace{S_0\ldots S_0}_{N-M}
\right).
\end{aligned}
\label{eq:triplet-averaged-population}
\end{equation}
This convention fixes the trace of every product population operator to unity,
\begin{equation}
\operatorname{Tr}
\left\{
\hat{P}\!\left(
\underbrace{T\ldots T}_{M}
\underbrace{S_0\ldots S_0}_{N-M}
\right)
\right\}
=
1,
\label{eq:population-trace-one}
\end{equation}
which is convenient for the following construction of the orthogonal basis set of long-lived operators. The Hilbert--Schmidt norms are not the same for different triplet numbers:
\begin{equation}
\left\|
\hat{P}(
\underbrace{T\ldots T}_{M}
\underbrace{S_0\ldots S_0}_{N-M}
)
\right\|
=
\frac{1}{\sqrt{3^M}}.
\label{eq:population-norm}
\end{equation}
Importantly, population operators with different arrangements of the $T$ and $S_0$ labels are mutually orthogonal.

It is therefore useful to organize the subspaces of LLOs in Eq.~(\ref{eq:many-pair-product-operator}) according to the number of triplet labels in the operator, as shown in Fig.~\ref{fig:1}c. We denote by $P(MT)$ the subspace spanned by all population operators containing exactly $M$ triplet labels and $N-M$ singlet labels:
\begin{equation}
\begin{aligned}
P(MT)
&=
\operatorname{span}
\left\{
\begin{array}{@{}c@{}}
\hat{P}\!\left(
\underbrace{T\ldots T}_{M}
\underbrace{S_0\ldots S_0}_{N-M}
\right)
\\[0.3em]
\text{and all distinct label permutations}
\end{array}
\right\}.
\end{aligned}
\label{eq:PMT-definition}
\end{equation}
For example, $P(0T)$ contains the unique all-singlet configuration $\hat{P}(S_0S_0\ldots S_0)$, while $P(NT)$ contains the unique all-triplet configuration $\hat{P}(TT\ldots T)$. In general, the number of distinct basis operators in $P(MT)$ is the number of ways to choose the positions of the $M$ triplet groups among $N$ sites,
\begin{equation}
\dim\left\{P(MT)\right\}
=
\binom{N}{M}.
\label{eq:dim-PMT}
\end{equation}

The full dimension of the zero-eigenvalue space generated by the intra-pair relaxation mechanism is therefore equal to
\begin{equation}
\sum_{M=0}^{N}
\text{dim}\left\{P(MT)\right\}
=
\sum_{M=0}^{N}
\binom{N}{M}
=
2^N.
\label{eq:dim-full-population-space}
\end{equation}
One of these $2^N$ operators is always the trivial conserved population of all states, which is given by the identity operator. Consequently, the number of non-trivial long-lived operators is
\begin{equation}
N_{\mathrm{LLO}}
=
2^N-1.
\label{eq:number-LLS}
\end{equation}

Although the individual populations operators of the singlet--triplet states form a convenient basis for the zero-eigenvalue subspace, they are not the most useful physical basis. Experimentally, one excites population imbalances rather than the total population itself. Therefore, the relevant long-lived operators are linear combinations of the population operators. 

For an $AA'MM'XX'\ldots$ spin system, as relevant to aliphatic chains in achiral molecules, only operators that connect states with the same global intra-pair permutation parity are directly accessible. Since a singlet state is antisymmetric with respect to intra-pair spin permutation, whereas the triplet manifold is symmetric, a product state containing $M$ triplet groups and $N-M$ singlet groups has global intra-pair parity
$(-1)^{N-M}$. Thus, the spaces $P(MT)$ naturally split into two manifolds according to this parity. This motivates the construction of a symmetry-adapted basis within each $P(MT)$ subspace, which is carried out below using the permutation symmetry of the $N$ spin pairs.

\subsubsection{Irreducible Representations of the Permutation Group}
Strictly speaking, the subspace \(P(MT)\) carries the representation of the permutation group \(S_N\) acting on the \(N\) spin pairs. This action simply permutes the positions of the \(M\) triplet labels among the \(N\) available positions. This representation is reducible and can be decomposed into irreducible representations of \(S_N\). These irreducible representations are labelled by Young diagrams. \cite{hamermesh1962group,messiah1962quantum,landau1977quantum} A Young diagram corresponding to a partition of \(N\) is denoted by
$
[\lambda_1,\lambda_2,\ldots,\lambda_M],
$
where
$
\lambda_1+\lambda_2+\cdots+\lambda_M=N,
\qquad
\lambda_1\geq \lambda_2\geq \cdots \geq \lambda_M.
$ The diagram consists of \(M\) rows with $\lambda_1,\lambda_2,\ldots,\lambda_M$ boxes, respectively, as explained below.

Since every position can carry only one of two labels, \(S_0\) or \(T\), the relevant Young diagrams can contain only one or two rows, therefore, only Young diagrams of the form
$[N-r,r]$
can appear. This is the standard decomposition of the permutation representation of $S_N$ acting on $M$-element subsets. 

The decomposition of the ${P}(MT)$ subspace can be written as
\begin{equation}
{P}(MT)
=
\bigoplus_{r=0}^{\min(M,N-M)}
[N-r,r].
\label{eq:decompostition}
\end{equation}
The upper limit \(\min(M,N-M)\) has a simple origin. First, the Young diagram \([N-r,r]\) must be a valid two-row diagram, so the second row cannot be longer than the first one. Second, the second row cannot contain more boxes than the number of labels of either type, because the representation is generated by configurations with \(M\) triplet labels and \(N-M\) singlet labels. Hence,
\begin{equation}
r=0,1,\ldots,\min(M,N-M).
\end{equation}

Thus, \(M\) specifies the population subspace, i.e. how many spin pairs are in triplet states, whereas \(r\) specifies how this component transforms under permutations of the \(N\) spin pairs. The component with \(r=0\) corresponds to the fully symmetric combination within $P(MT)$, whereas components with larger \(r\) describe less symmetric modes,

For example, for \(N=4\) and \(M=2\), there are only three relevant Young diagrams
\begin{equation}
[4,0], \qquad [3,1], \qquad [2,2].
\end{equation}
They can be represented schematically as
\begin{equation}
[4,0]: \\ \yng(4)
\qquad
[3,1]: \\ \yng(3,1)
\qquad
[2,2]: \\ \yng(2,2)
\end{equation}
Each Young diagram labels a different irreducible permutation-symmetry component, and different components are mutually orthogonal.

\subsubsection{The Construction of the Irreducible Basis Operators}

The irreducible basis set can be constructed using Young symmetrizers. For a fixed value of \(M\), the permutation representation carried by \(P(MT)\) is given in Eq.(\ref{eq:decompostition}), where $[N-r,r]$ denotes a Young diagram, i.e. only the shape of the irreducible representation. For each shape, one should in general consider the corresponding Young tableaux, obtained by filling the boxes of the diagram with the labels \(1,\ldots,N\). Different tableaux of the same shape define different Young symmetrizers and may generate different basis operators within the same irreducible symmetry sector. This is explained in detail below.

The construction proceeds as follows:
\begin{enumerate}
    \item Fix \(N\) and \(M\), and choose an allowed Young diagram
    \begin{equation}
        [N-r,r],
        \qquad
        r=0,1,\ldots,\min(M,N-M).
    \end{equation}

    \item For this Young diagram, construct the required Young tableaux $t$ by filling the boxes with the indices \(1,\ldots,N\). For example, for the shape \([3,1]\) that correspinds to $N=4$, $r=1$, possible tableaux include
    \begin{equation}
    \begin{array}{ccc}
    1 & 2 & 3 \\
    4
    \end{array}
    \qquad
    \begin{array}{ccc}
    1 & 2 & 4 \\
    3
    \end{array}
    \qquad
    \begin{array}{ccc}
    1 & 3 & 4 \\
    2
    \end{array}.
    \label{eq:Young}
    \end{equation}
Here, the entries must increase from left to right along each row and from top to bottom along each column. These ordering conditions exclude, for example, the filling with $2,3,4$ in the first row and $1$ in the second row, since the entries in the first column would not increase from top to bottom.
    \item For each tableau \(t\), construct the row symmetrization superoperator \(\hat{\hat{S}}_{t}\), which symmetrizes over all indices belonging to the same row of \(t\). This is constructed via summation of all possible permutations of the indices in each row. For example, for the first tableau of Eq.(\ref{eq:Young}) this gives the following symmetrization superoperator for the first row: $\hat{\hat{S}}_{t,1}=\hat{\hat{E}}+\hat{\hat{\Pi}}_{12}+\hat{\hat{\Pi}}_{23}+\hat{\hat{\Pi}}_{13}+\hat{\hat{\Pi}}_{12}\hat{\hat{\Pi}}_{23}+\hat{\hat{\Pi}}_{12}\hat{\hat{\Pi}}_{13}$, and for the second row there is only trivial permutation: $\hat{\hat{S}}_{t,2}=\hat{\hat{E}}$. The total symmetrization superoperator is given by $\hat{\hat{S}}_{t}=\hat{\hat{S}}_{t,2}\hat{\hat{S}}_{t,1}$. The superoperator $\hat{\hat{\Pi}}_{ij}$ permutes the labels of the $-\mathrm{CH}_2-$ groups with indices $i$ and $j$. For instance,
$
\hat{\hat{\Pi}}_{13}\hat{P}(TS_0S_0S_0)=\hat{P}(S_0S_0TS_0).
$

    \item For the same tableau $t$, we construct the column antisymmetrization superoperator $\hat{\hat{A}}_{t}$. This operator antisymmetrizes over all indices belonging to the same column of $t$. In general, for a given column $C$, the corresponding antisymmetrizer is obtained by summing over all permutations $\pi\in S_C$ of the indices inside this column with the factor $\mathrm{sgn}(\pi)$:
$
\hat{\hat{A}}_{C}=\sum_{\pi\in S_C}\mathrm{sgn}(\pi)\hat{\hat{\Pi}}_{\pi}.
$
Here, $\mathrm{sgn}(\pi)=+1$ for an even permutation and $\mathrm{sgn}(\pi)=-1$ for an odd permutation. The full column antisymmetrizer is then the product over all columns,
$
\hat{\hat{A}}_{t}=\prod_{C\in t}\hat{\hat{A}}_{C}.
$
For the first tableau of Eq.~(\ref{eq:Young})
there is only one non-trivial column, $C_1={1,4}$. Hence,
$
\hat{\hat{A}}_{C_1}=\hat{\hat{E}}-\hat{\hat{\Pi}}_{14}.
$
The other columns, $C_2={2}$ and $C_3={3}$, contain only one index and therefore contribute only $\hat{\hat{E}}$. Thus,
$
\hat{\hat{A}}_{t}=\hat{\hat{E}}-\hat{\hat{\Pi}}_{14}.
$

    \item Construct the corresponding Young superoperator
    \begin{equation}
        \hat{\hat{Y}}_{t}
        =
        \hat{\hat{A}}_{t}
        \hat{\hat{S}}_{t}.
    \label{eq:Young}
    \end{equation}

    \item Apply $\hat{\hat{Y}}_{t}$ to the population operators spanning \(P(MT)\).

    \item Repeat this procedure for the required tableaux \(t\) of the same shape \([N-r,r]\). Remove all zero operators and linearly dependent combinations. The remaining operators span the component of symmetry type \([N-r,r]\) within \(P(MT)\).

    \item Repeat the construction for all allowed values of \(r\). The resulted operators belonging to different Young diagrams \([N-r,r]\) are mutually orthogonal. Within each irreducible representation, the obtained operators can be further orthogonalized, for example by the Gram--Schmidt procedure. \cite{golub2013matrix}

    \item The union of all symmetry-adapted operators obtained for all $r=0,1,\ldots,\min(M,N-M)$
    forms an orthogonal basis of $P(MT)$.

    \item The same construction is then repeated independently for all triplet-number sectors \(P(MT)\), with $M=0,1,\ldots,N$.
\end{enumerate}

The result of this construction is a permutation-adapted operator basis of the form (Fig.~\ref{fig:1}d)
\begin{equation}
\begin{aligned}
&\mathcal{B}(MT)
=
\\
&\left\{
\begin{array}{@{}l@{}}
\hat{B}_{r,\alpha}(MT):
\quad
r=0,1,\ldots,\min(M,N-M),
\\
\alpha=1,\ldots,d_r
\end{array}
\right\}.
\end{aligned}
\label{eq:B-basis-MT}
\end{equation}
where \(r\) labels the corresponding Young diagram \([N-r,r]\), while \(\alpha\) labels different basis operators within the same diagram. The number of operators in the diargam \([N-r,r]\) is well-known from the group theory and is equal to
\begin{equation}
d_r
=
\dim [N-r,r]
=
\binom{N}{r}
-
\binom{N}{r-1},
\qquad
\binom{N}{-1}\equiv 0.
\end{equation}
Thus, the total number of basis operators is
\begin{equation}
\sum_{r=0}^{\min(M,N-M)} d_r
=
\binom{N}{M},
\end{equation}
as required.

For example, the first few triplet manifolds have the following structure:
\begin{equation}
\begin{aligned}
P(0T):\qquad
& \hat{B}_{0,1}(0T),\\
P(1T):\qquad
& \hat{B}_{0,1}(1T),
\quad
\hat{B}_{1,\alpha}(1T),
\quad
\alpha=1,\ldots,N-1,\\
P(2T):\qquad
& \hat{B}_{0,1}(2T),
\quad
\hat{B}_{1,\alpha}(2T),
\quad
\alpha=1,\ldots,N-1,\\
& \hat{B}_{2,\beta}(2T),
\quad
\beta=1,\ldots,\frac{N(N-3)}{2},
\end{aligned}
\end{equation}
which is also shown in Fig.~\ref{fig:1}e for $N=4$, $M=2$. Here \(\hat{B}_{0,1}(MT)\) is the fully symmetric operator in the subspace \(P(MT)\). For simplicity, we will often denote it as \(\hat{B}_0(MT)\). The fully symmetric matrix representations of the operators \(\hat{B}_0(MT)\) are normalized to have the unit trace:
\begin{equation}
\operatorname{Tr}\left\{\hat{B}_0(MT)\right\}=1.
\end{equation}
Since the operators with $r\neq0$ are orthogonal to that with $r=0$ within the fixed $P(MT)$ subspace, this ensures that all other operators are traceless:
\begin{equation}
\operatorname{Tr}\left\{\hat{B}_{r,\alpha}(MT)\right\}=0,
\qquad r\neq 0.
\end{equation}
Therefore, the operators \(\hat{B}_{r,\alpha}(MT)\) with \(r\neq 0\) already represent traceless LLOs, that are linearly independent of the unity operator (a trivial zero-eigenvalue operator).

\subsection{The Triplet-Number Subspaces and Delocalized Eigenstates of Idealized Aliphatic Chains}

\begin{figure*}[t]
\centering
\includegraphics[width=16cm]{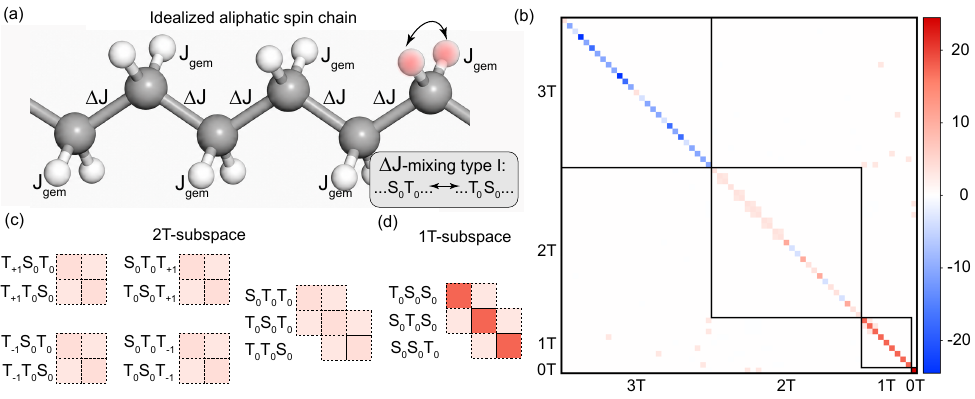}
\caption{
(a) Fragment of an aliphatic chain highlighting the two most important interactions in the system: $J_{\mathrm{gem}}$ denotes the geminal intra-pair coupling, whereas $\Delta J$ denotes the difference between the vicinal couplings to neighboring groups. In an idealized chain, both $J_{\mathrm{gem}}$ and $\Delta J$ are assumed to be uniform along the chain. (b) The matrix elements of the Hamiltonian $\frac{H_J}{2\pi}$ in the localized singlet--triplet basis, where the $\Delta J$--term mixes different states. Two types of mixing are present: type I mixing conserves the number of triplet labels in the wavefunction, whereas type II mixing changes it by two. Type II mixing is suppressed when $|J_{\mathrm{gem}}| \gg |\Delta J|$. In this limit, the blocks with different numbers of triplet labels, indicated by bold lines, can be treated independently. This ensures $P(MT)$ manifolds used for the construction of the LLS can also be treated independently. (c), (d) Type I mixing leads to delocalization, i.e., mixing of singlet--triplet wavefunctions between neighboring $-\mathrm{CH}_2-$ groups. For $N=3$, the $2T$ subspace contains five distinct mixed sub-blocks (c), whereas the $1T$ subspace contains only one such sub-block (d).
}
\label{fig:2}
\end{figure*}

Before constructing the explicit long-lived operators, it is useful to clarify the role of the triplet-number subspaces $P(MT)$. These subspaces are not only a convenient combinatorial classification of the LLOs, but also have a direct physical meaning for aliphatic spin chains in achiral molecules. The coherent part of $J$-coupling Hamiltonian of an "idealized" aliphatic spin chain  \cite{sheberstov2025aliphatic} is given by (see also Fig.~\ref{fig:2}a):

\begin{equation}
\begin{aligned}
&\hat{H}_{J} = 2\pi J_{\mathrm{gem}} \sum_{k=1}^{N}
\hat{\mathbf{I}}^{(2k-1)}\cdot\hat{\mathbf{I}}^{(2k)}
\\
&+2\pi\frac{\Sigma J}{2} \sum_{k=1}^{N-1}
\left( \hat{I}^{(2k-1)}_z+\hat{I}^{(2k)}_z \right)
\left( \hat{I}^{(2k+1)}_z+\hat{I}^{(2k+2)}_z \right)
\\
&+2\pi\frac{\Delta J}{2} \sum_{k=1}^{N-1}
\left( \hat{I}^{(2k-1)}_z-\hat{I}^{(2k)}_z \right)
\left( \hat{I}^{(2k+1)}_z-\hat{I}^{(2k+2)}_z \right).
\end{aligned}
\label{eq:H-J-aliphatic-chain}
\end{equation}

Here, $J_{\mathrm{gem}}=J^{(1,2)}=J^{(3,4)}=\cdots$ is the geminal $J$-coupling between the two protons attached to the same carbon atom, and $N$ denotes the total number of $-\mathrm{CH}_2-$ groups in the aliphatic chain. The quantities $\Sigma J$ and $\Delta J$ are the sum and difference of the two vicinal couplings between neighboring $-\mathrm{CH}_2-$ groups,
$
\Sigma J = J_{\mathrm{gauche}}+J_{\mathrm{anti}},
\Delta J = J_{\mathrm{gauche}}-J_{\mathrm{anti}}.
$
The operator $\hat{\mathbf{I}}^{(k)}$ is the vector spin operator of spin $i$, $\hat{\mathbf{I}}^{(k)}=\{\hat{I}^{(k)}_x,\hat{I}^{(k)}_y,\hat{I}^{(k)}_z\}$.

The $\Delta J$--term of Eq.~(\ref{eq:H-J-aliphatic-chain}) is responsible for delocalization, i.e., for coherent mixing of localized singlet--triplet wavefunctions belonging to neighboring $-\mathrm{CH}_2-$ groups. \cite{sheberstov2024collective,sheberstov2025aliphatic} In general, this term can induce two types of mixing. The type I mixing conserves the number of triplet labels and mixes states of the form
$
|\ldots S_0 T_0 \ldots\rangle
\leftrightarrow
|\ldots T_0 S_0 \ldots\rangle .
$
The type II mixing changes the number of triplet labels by two and couples states of the form
$
|\ldots S_0 S_0 \ldots\rangle
\leftrightarrow
|\ldots T_0 T_0 \ldots\rangle .
$

For aliphatic spin chains in achiral molecules, the intra-pair scalar coupling is the largest coherent interaction in the system,
$
|J_{\mathrm{gem}}|\gg|\Delta J|.
\label{eq:Jgem-large}
$
In this limit, the mixing of the second type is strongly suppressed. As a result, the number of triplet labels is approximately conserved by the coherent dynamics, and the subspaces
$
P(0T),\ P(1T),\ \ldots,\ P(NT)
$
can be treated as independent blocks of the spin-Hamiltonian given in Eq.~\eqref{eq:H-J-aliphatic-chain}, as also shown in Fig.~\ref{fig:2}b, where mixing between wavefunctions with different numbers of triplet labels is suppressed. Within each fixed $P(MT)$ block, however, the $\Delta J$--term still mixes the localized states with the same number of triplet labels, giving rise to delocalized eigenstates. We also note that the present analysis is focused on achiral molecules, for which the two protons within each methylene group have identical chemical shifts, and only the $J$--coupling Hamiltonian of Eq.~\eqref{eq:H-J-aliphatic-chain} is relevant. In chiral molecules, the corresponding intra-pair chemical-shift differences must be included in the spin Hamiltonian, and these additional terms modify the delocalized eigenstates and, in general, lead to a different eigenbasis. A detailed analysis of long-lived operators in the delocalized eigenbasis relevant to chiral molecules is beyond the scope of the present work.

Returning to the achiral molecules considered in this work, we illustrate the emergence of delocalized eigenstates using the localized states of the $P(1T)$ subspace:
\begin{equation}
\nonumber
T_0S_0S_0\ldots S_0,\quad
S_0T_0S_0\ldots S_0,\quad
\ldots,\quad
S_0S_0\ldots S_0T_0.
\end{equation}
These localized states are not eigenstates of the coherent Hamiltonian, \cite{sheberstov2025aliphatic} as shown in Fig.~\ref{fig:2}d for $N=3$ methylene groups. Instead, they are mixed by the $\Delta J$--term and form delocalized linear combinations. Similarly, within the $P(2T)$ subspace, the localized states with two triplet labels mix with one another to form delocalized eigenstates, as illustrated in Fig.~\ref{fig:2}c for $N=3$ methylene groups. Since the different $P(MT)$ manifolds remain separated, it is convenient to treat the states within each manifold using a permutation-adapted basis.

This observation explains why a permutation-adapted basis within each $P(MT)$ subspace is particularly useful. The operator $\hat{B}_0(MT)$ is fully symmetric with respect to permutations of the $N$ spin pairs and therefore represents the invariant population component of the corresponding manifold. As a result, linear combinations of such invariant operators contain only populations in the delocalized basis set. They do not generate coherences when transformed to the delocalized eigenbasis of the coherent Hamiltonian, as shown below.

In contrast, the operators $\hat{B}_{r,\alpha}(MT)$ with $r\neq0$ belong to non-trivial irreducible representations of the permutation group within a fixed $P(MT)$ manifold. Since the localized states inside this manifold are coherently mixed by the $\Delta J$--term, these non-invariant population combinations generally transform into operators containing both populations and coherences in the delocalized eigenbasis. This provides the origin of the collective long-lived coherences associated with the $\hat{B}_{r,\alpha}(MT)$ operators for $r\neq0$. The basis inside each irreducible component is not unique; for convenience, we choose it to be adapted to the reflection symmetry of the chain, so that the resulting operators are either symmetric or antisymmetric under reversal of the $-\mathrm{CH}_2-$ order. This additional convention only fixes the basis within a given irreducible subspace and should not be confused with the global intra-pair $g/u$ parity discussed above.

\enlargethispage{3\baselineskip}

Thus, the construction above naturally separates the long-lived operators into two classes. The first class is formed from the invariant population combinations $\hat{B}_0(MT)$. The linear combinations of these operators give purely long-lived population imbalances, i.e., LLSs. The second class consists of the non-invariant operators $\hat{B}_{r,\alpha}(MT)$ with $r\neq0$, which belong to non-trivial irreducible representations within fixed $P(MT)$ manifolds. In the localized singlet--triplet basis, these operators appear as population imbalances; however, in the delocalized eigenbasis they generally contain both long-lived population imbalances and long-lived coherences. This is shown in the following sections.

\subsection{The Construction of the Delocalized Long-Lived States and Collective Zero-Quantum Coherences}

The operators \(\hat{B}_{r,\alpha}(MT)\) with \(r\neq 0\) can already be used as long-lived operators since they are orthogonal and traceless, see Fig.~\ref{fig:1}f, and are therefore linearly independent of the identity operator (a trivial zero-eigenvalue mode that is always present for any spin system). As we show below, these operators always contain a combination of populations and coherences, therefore giving information on both the LLSs and LLCs.

The operators \(\hat{B}_0(MT)\), in contrast, cannot be used directly as LLOs because each of them has a nonzero trace, making them linearly dependent with an identity operator. The appropriate, traceless LLOs should therefore be constructed as differences between permutation-symmetric population operators \(\hat{B}_0(MT)\) belonging to triplet-number manifolds with the same global intra--pair permutation parity, since only these imbalances can be experimentally excited. This is because the global intra--pair parity of the aliphatic chain in achiral molecules is conserved, and the $g/u$--manifolds are independent of each other. As we show below, in case of achiral molecules, these combinations of  \(\hat{B}_0(MT)\) do not contain coherences, and therefore they represent LLSs separated from coherences. This algorithm of construction of the LLSs and LLCs is schematically shown in Fig.~\ref{fig:1}f for $N=4$ methylene groups.

The global intra-pair parity of the states within the subspace \(P(MT)\) is equal to $(-1)^{N-M}$.
Hence the subspaces
\begin{equation}
\nonumber
P(NT),\ P((N-2)T),\ P((N-4)T),\ldots
\end{equation}
contain symmetric (with respect to the global parity) states, whereas the subspaces
\begin{equation}
\nonumber
P((N-1)T),\ P((N-3)T),\ P((N-5)T),\ldots
\end{equation}
contain antisymmetric states. The symmetry is denoted via the superscript: $g$ corresponds to the LLOs involving $gerade$ states, whereas $u$ denotes LLOs involving $ungerade$ states.

Consequently, the traceless permutation-invariant LLSs can be constructed separately within the two $g/u$--manifolds. In the $g$--manifold, we start from the set of symmetric operators
$
\hat{B}_0(NT),\hat{B}_0((N-2)T),\hat{B}_0((N-4)T),\ldots,
$
whereas in the $u$--manifold we use
$
\hat{B}_0((N-1)T),\hat{B}_0((N-3)T),\hat{B}_0((N-5)T),\ldots.
$
Only terms with non-negative triplet numbers are retained. Within each manifold, traceless mutually orthogonal LLS operators are then obtained by applying the Gram--Schmidt procedure with the additional requirement that each resulting operator is orthogonal to the identity operator. Equivalently, each constructed operator must satisfy
$
\mathrm{Tr}\left\{\hat{L}\right\}=~0.
$
The resulting operators are not necessarily normalized, but they can always be divided by their Hilbert--Schmidt norms.

The number of such independent traceless LLSs is equal to the number of subspaces in each $g/u$--manifold minus one. Thus, for the $g$--manifold containing \(P(NT),P((N-2)T),\ldots\), the number of the LLSs is
\begin{equation}
n^{(g)}
=
\left\lfloor\frac{N}{2}\right\rfloor.
\end{equation}
For the $u$--manifold, containing \(P((N-1)T),P((N-3)T),\ldots\), the number of LLSs is
\begin{equation}
n^{(u)}
=
\left\lfloor\frac{N-1}{2}\right\rfloor.
\end{equation}
Thus, if \(N\) is odd, the two manifolds contain the same number of permutation-symmetric LLS operators,
\begin{equation}
n^{(g)}=n^{(u)}=\frac{N-1}{2}.
\end{equation}
If \(N\) is even, the $g$--manifold contains one more subspace, and therefore
\begin{equation}
n^{(g)}=\frac{N}{2},
\qquad
n^{(u)}=\frac{N}{2}-1.
\end{equation}

Finally, together with \(\hat{B}_{r,\alpha}(MT)\), \(r\neq 0\), the total number of traceless LLOs generated by the construction described above is
$2^N-2$. This is because there are always $2^{N-1}-1$ traceless operators within the $g$--manifold, which consists of $P(NT)$, $P((N-2)T)$, $P((N-4)T)$, $\ldots$, and the same number of traceless operators within the $u$--manifold, which consists of $P((N-1)T)$, $P((N-3)T)$, $P((N-5)T)$, $\ldots$. This counting can be shown formally using elementary combinatorics via the summation of Eq.~(\ref{eq:dim-PMT}) over odd and even $M$ separately.

As discussed in the previous section, the total number of LLOs is $2^N-1$. Therefore, the present construction provides $2^N-2$ operators, meaning that the one LLO is always missing. This remaining LLO necessarily involves a population imbalance between the states that have different parity. Consequently, this operator cannot be excited in experimentally relevant cases of achiral molecules since the global parity is always conserved. For this reason, one can excite no more than $2^N-2$ LLOs in achiral molecules but up to $2^N-1$ LLOs in chiral molecules. For completeness, for $N=2,3,4$ we also provide the explicit form of the remaining long-lived operator, $\hat{L}_r$, that can be excited only in chiral molecules. This remaining operator $\hat{L}_r$ always has the following structure:
\begin{equation}
    \hat{L}_r=C\left(\frac{\mathbf{\hat{1}_u}}{\text{Tr}\{\mathbf{\hat{1}_u\}}}-\frac{\mathbf{\hat{1}_g}}{\text{Tr}\mathbf{\{\hat{1}_g\}}}\right).
\label{eq:Lr}
\end{equation}

Here, $\hat{\mathbf{1}}_g$ and $\hat{\mathbf{1}}_u$ are the identity operators acting in the $g$- and $u$-parity subspaces, respectively. The resulting operator is traceless, and $C$ is a normalization factor chosen such that $\|\hat{L}_r\|=1$.

\section{Results and Discussion}

\subsection{The Construction of the Long-Lived Operators for $-(\text{CH}_2)_2-$}

For $N=2$, there are three subspaces: $P(0T)$, $P(1T)$, and $P(2T)$. The subspaces $P(0T)$ and $P(2T)$ are one-dimensional:
\begin{equation}
P(0T)=\operatorname{span}\left\{\hat{P}(S_0S_0)\right\},\\
P(2T)=\operatorname{span}\left\{\hat{P}(TT)\right\}.
\end{equation}
Both of these sectors contain only the fully symmetric Young component \([2,0]\).

The subspace \(P(1T)\) is two-dimensional:
\begin{equation}
P(1T)
=
\operatorname{span}
\left\{
\hat{P}(TS_0),\hat{P}(S_0T)
\right\}.
\end{equation}
For \(M=1\), the allowed Young diagrams are
[2,0],
[1,1]. Let us consider each diagram.

\subsubsection{The Young diagram $\mathbf{[2,0]}$}

The symmetric ($r=0$) basis operator can be obtained from the only allowed tableau
\begin{equation}
t=
\begin{array}{cc}
1 & 2
\end{array}.
\end{equation}
For this tableau, both indices belong to the same row. Therefore, there are only two permutations, and the row symmetrization superoperator introduced in Eq.~\eqref{eq:Young} is
\begin{equation}
\hat{\hat{S}}_t
=
\hat{\hat{E}}
+
\hat{\hat{\Pi}}_{12}.
\end{equation}
There is no non-trivial column antisymmetrization, and hence
\begin{equation}
\hat{\hat{A}}_t=\hat{\hat{E}}.
\end{equation}
The corresponding Young superoperator is
\begin{equation}
\hat{\hat{Y}}_t
=
\hat{\hat{A}}_t\hat{\hat{S}}_t
=
\hat{\hat{E}}
+
\hat{\hat{\Pi}}_{12}.
\end{equation}
Applying this superoperator to \(\hat{P}(TS_0)\), we obtain
\begin{equation}
\hat{\hat{Y}}_t\hat{P}(TS_0)
=
\hat{P}(TS_0)+\hat{P}(S_0T).
\end{equation}
After normalization, the symmetric basis operator is
\begin{equation}
\hat{B}_{0}(1T)
=
\frac{1}{2}
\left(
\hat{P}(TS_0)+\hat{P}(S_0T)
\right).
\end{equation}
The normalization ensures $Tr\left\{\hat{B}_{0}(1T)\right\}=1$. This operator belongs to the fully symmetric irreducible representation \([2,0]\) of the subspace $P(1T)$.

\subsubsection{The Young diagram $\mathbf{[1,1]}$}

The only Young tableau for the antisymmetric component \([1,1]\) of $P(1T)$ is
\begin{equation}
\begin{array}{c}
1\\
2
\end{array}.
\end{equation}

The row symmetrization superoperator \(\hat{\hat{S}}_t\) is constructed by summing all permutations of the indices that belong to the same row, with positive signs. In the present case, each row contains only one index. Therefore, the row symmetrizer is simply the identity superoperator,
\begin{equation}
\hat{\hat{S}}_t=\hat{\hat{E}}.
\end{equation}

The column antisymmetrization superoperator \(\hat{\hat{A}}_t\), see Eq.~\eqref{eq:Young}, is constructed by summing all permutations of the indices that belong to the same column, with the sign of the corresponding permutation. Since the tableau has one column containing the two indices \(1\) and \(2\), we obtain
\begin{equation}
\hat{\hat{A}}_t
=
\hat{\hat{E}}
-
\hat{\hat{\Pi}}_{12},
\end{equation}

The corresponding Young superoperator is therefore
\begin{equation}
\hat{\hat{Y}}_t
=
\hat{\hat{A}}_t\hat{\hat{S}}_t
=
\hat{\hat{E}}
-
\hat{\hat{\Pi}}_{12}.
\end{equation}
Applying this operator to \(\hat{P}(TS_0)\) gives
\begin{equation}
\hat{\hat{Y}}_t \hat{P}(TS_0)
=
\hat{P}(TS_0)-\hat{P}(S_0T).
\end{equation}
This gives the antisymmetric traceless basis operator $\hat{B}_{1,1}(1T)$.

\subsubsection{The Long-Lived Operators}

Thus, for \(N=2\), the full basis is
\begin{equation}
\begin{aligned}
P(0T):\qquad
& \hat{B}_0(0T)=\hat{P}(S_0S_0),
\\
P(1T):\qquad
& \hat{B}_0(1T)
=
\frac{1}{2}
\left(
\hat{P}(TS_0)+\hat{P}(S_0T)
\right),\\
& \hat{B}_{1,1}(1T)
=
\hat{P}(TS_0)-\hat{P}(S_0T),
\\
P(2T):\qquad
& \hat{B}_0(2T)=\hat{P}(TT).
\end{aligned}
\end{equation}

The traces are
\begin{equation}
\operatorname{Tr}\{\hat{B}_0(0T)\}
=
\operatorname{Tr}\{\hat{B}_0(1T)\}
=
\operatorname{Tr}\{\hat{B}_0(2T)\}
=
1,
\end{equation}
whereas
\begin{equation}
\operatorname{Tr}\{\hat{B}_{1,1}(1T)\}=0.
\end{equation}
Therefore, \(\hat{B}_{1,1}(1T)\) is already a traceless imbalance operator corresponding to the $u$--manifold.

The fully symmetric operators \(\hat{B}_0(0T)\) and \(\hat{B}_0(2T)\) correspond to states the same global intra-pair parity, $g$, because their parity factors are equal to +1. Hence the Gram--Schmidt procedure in this case is trivial since they can be combined into the traceless LLO
\begin{equation}
\begin{aligned}
&\hat{L}^{(g)}_0
=
\hat{B}_0(0T)-\hat{B}_0(2T)
=
\hat{P}(S_0S_0)-\hat{P}(TT).
\end{aligned}
\end{equation}

The LLO in the $u$--manifold is given by $\hat{B}_{1,1}(1T)$:
\begin{equation}
\begin{aligned}
\hat{L}^{(u)}_0
=
\hat{B}_{1,1}(1T)
=
\hat{P}(TS_0)-\hat{P}(S_0T).
\end{aligned}
\end{equation}

The remaining LLO describes a population imbalance between the
subspaces of different parity, $g$ and $u$. The identity operators
restricted to these subspaces can be written as
\begin{equation}
\mathbf{\hat{1}_g}
=
\hat{P}(S_0S_0)
+
9\hat{P}(TT), 
\quad\mathbf{\hat{1}_u}
=
3\hat{P}(S_0T)
+
3\hat{P}(TS_0).
\end{equation}
Using Eq.~\eqref{eq:Lr}, the corresponding remaining operator is given by
\begin{equation}
\begin{aligned}
&\hat{L}_r
=
C
\left(
\frac{1}{6}\mathbf{\hat{1}_u}-\frac{1}{10}\mathbf{\hat{1}_g}
\right)
\\
&=
\tilde{C}\Bigl(
5\hat{P}(TS_0)
+5\hat{P}(S_0T)
-\hat{P}(S_0S_0)
-9\hat{P}(TT)
\Bigr).
\end{aligned}
\label{eq:remaining}
\end{equation}
We emphasize that an LLO must be traceless in order to be orthogonal to the trivial conserved operator, namely the identity. Thus, the total number of non-trivial LLOs is $2^N-1$, which gives three linearly independent operators for $N=2$. Thus, the final orthonormal basis set is given by division of the LLOs by their norms, i.e., each operator is normalized as $||\hat{L}||=1$.

\begin{equation}
\begin{aligned}
&\hat{L}^{(g)}_0
=
\frac{3}{\sqrt{10}}
\Bigl\{
\hat{P}(S_0S_0)-\hat{P}(TT)
\Bigr\},
\\[0.3em]
&\hat{L}^{(u)}_0
=
\sqrt{\frac{3}{2}}
\Bigl\{
\hat{P}(TS_0)-\hat{P}(S_0T)
\Bigr\},
\\[0.3em]
&\hat{L}_r
=
\frac{1}{4}
\sqrt{\frac{3}{5}}
\Bigl(
5\hat{P}(TS_0)
+5\hat{P}(S_0T)
\\[-0.2em]
&\quad
-\hat{P}(S_0S_0)
-9\hat{P}(TT)
\Bigr).
\end{aligned}
\label{eq:LLS-N2}
\end{equation}

\begin{table}[t]
\caption{
Expansion coefficients of the long-lived operators in the $g/u$--manifolds for $N=2$ in the scalar-product-operator basis set
$\hat{X}^{(1,2)}=\hat{\mathbf I}^{(1)}\cdot\hat{\mathbf I}^{(2)}$ and
$\hat{X}^{(3,4)}=\hat{\mathbf I}^{(3)}\cdot\hat{\mathbf I}^{(4)}$. The indices $(1,2)$ and $(3,4)$ correspond to the first and second $-\mathrm{CH}_2-$ groups, respectively. All operators are normalized to unit Hilbert--Schmidt norm, i.e., $||\hat{L}||=1$.
}
\centering
\renewcommand{\arraystretch}{3}
\setlength{\tabcolsep}{8pt}
\begin{tabular}{c c c c}
\hline
Operator
&
$\hat{X}^{(1,2)}$
&
$\hat{X}^{(3,4)}$
&
$\hat{X}^{(1,2)}\hat{X}^{(3,4)}$
\\
\hline
$\hat{L}^{(g)}_0$
&
$-\dfrac{1}{\sqrt{10}}$
&
$-\dfrac{1}{\sqrt{10}}$
&
$\dfrac{4\sqrt{10}}{15}$
\\
$\hat{L}^{(u)}_0$
&
$\dfrac{\sqrt{6}}{6}$
&
$-\dfrac{\sqrt{6}}{6}$
&
$0$
\\
\hline
\end{tabular}
\label{tab:N2-LLS-product-operator-coefficients}
\end{table}

First, we note that $\hat{L}_r$ cannot be excited in an $AA'XX'$ spin system (this corresponds to a spin chain in an achiral molecule) by any pulse sequence, since it involves a population imbalance between states of different global intra-pair parity, see Eqs.~\eqref{eq:Lr},~\eqref{eq:remaining}. The operator $\hat{L}^{(g)}_0$ represents a purely long-lived state (LLS). The operator $\hat{L}^{(u)}_0$, however, has a different structure. Although it is written as a population imbalance in the localized singlet--triplet basis, it contains both populations and coherences when expressed in the delocalized basis.

This follows from the fact that the $\Delta J$--term of the Hamiltonian mixes the localized states $T_0S_0$ and $S_0T_0$, forming the delocalized states \cite{sheberstov2025aliphatic}
\begin{align}
|\psi_1^{\text{deloc}}\rangle
&=
\frac{
|T_0S_0\rangle+|S_0T_0\rangle
}{\sqrt{2}}
,
\nonumber\\
|\psi_2^{\text{deloc}}\rangle
&=
\frac{|T_0S_0\rangle-|S_0T_0\rangle}{\sqrt{2}}.
\label{eq:psi12-TS-ST}
\end{align}
Therefore,
\begin{align}
\hat{P}(T_0S_0)+\hat{P}(S_0T_0)
&=
\hat{P}(\psi_1^{\text{deloc}})+\hat{P}(\psi_2^{\text{deloc}}),
\nonumber\\
\hat{P}(T_0S_0)-\hat{P}(S_0T_0)
&=
|\psi_1^{\text{deloc}}\rangle\langle\psi_2^{\text{deloc}}|
+
|\psi_2^{\text{deloc}}\rangle\langle\psi_1^{\text{deloc}}|.
\label{eq:populations-T0S-ST0-psi}
\end{align}
Thus, the operator $\hat{L}^{(u)}_0$ can be written as
\begin{align}
&\hat{L}^{(u)}_0
=
\sqrt{\frac{1}{6}}
\bigg\{
\sum_{i=\pm1}
\left(
\hat{P}(T_iS_0)-\hat{P}(S_0T_i)
\right)
\nonumber\\
&+
|\psi_1^{\text{deloc}}\rangle\langle\psi_2^{\text{deloc}}|
+ 
|\psi_2^{\text{deloc}}\rangle\langle\psi_1^{\text{deloc}}|
\bigg\}.
\label{eq:Lu0-N2-deloc}
\end{align}
This expression shows explicitly that $\hat{L}^{(u)}_0$ contains population imbalances, $\hat{P}(T_{\pm1}S_0)-\hat{P}(S_0T_{\pm1})$, as well as coherences between the delocalized states $|\psi_1\rangle$ and $|\psi_2\rangle$ originating from the $\Delta J$--mixing of $T_0S_0/S_0T_0$ states. 

It is important to note that the LLS operator $\hat{L}^{(g)}_0$ has indeed been excited experimentally using the SLIC pulse sequence. \cite{sonnefeld2022long}. In the same work, it has been shown that $\hat{L}^{(u)}_0$, which contains both LLS and LLC components, cannot be excited by the SLIC pulse sequence in achiral molecules. This does not preclude its excitation by other methods; however, to the best of our knowledge, its experimental observation in an achiral molecule has not yet been reported. By contrast, in chiral molecules, SLIC excitation can access such states containing LLCs and induce coherent exchange between unequally populated singlet orders of the two CH$_2$ groups \cite{devience2016probing}.

\subsection{The Long-Lived Operators for $-(\text{CH}_2)_3-$}
For $N=3$, the full irreducible basis set of operators is given by (see Appendix B for details):
\begin{equation}
\begin{aligned}
&P(0T):
\\[-0.2em]
&\quad
\hat{B}_0(0T)=\hat{P}(S_0S_0S_0),
\\[0.4em]
&P(1T):
\\[-0.2em]
&\quad
\hat{B}_0(1T)=\frac{1}{3}
\left(
\hat{P}(TS_0S_0)+\hat{P}(S_0TS_0)+\hat{P}(S_0S_0T)
\right),
\\[-0.2em]
&\quad
\hat{B}_{1,1}(1T)
=
\hat{P}(TS_0S_0)-\hat{P}(S_0S_0T),
\\[-0.2em]
&\quad
\hat{B}_{1,2}(1T)
=
\hat{P}(TS_0S_0)-2\hat{P}(S_0TS_0)+\hat{P}(S_0S_0T),
\\[0.4em]
&P(2T):
\\[-0.2em]
&\quad
\hat{B}_0(2T)=\frac{1}{3}
\left(
\hat{P}(S_0TT)+\hat{P}(TS_0T)+\hat{P}(TTS_0)
\right),
\\[-0.2em]
&\quad
\hat{B}_{1,1}(2T)
=
\hat{P}(S_0TT)-\hat{P}(TTS_0),
\\[-0.2em]
&\quad
\hat{B}_{1,2}(2T)
=
\hat{P}(S_0TT)-2\hat{P}(TS_0T)+\hat{P}(TTS_0),
\\[0.4em]
&P(3T):
\\[-0.2em]
&\quad
\hat{B}_0(3T)=\hat{P}(TTT).
\end{aligned}
\label{eq:B-basis-N3}
\end{equation}

The traceless operators $\hat{B}_{1,\alpha}$ already represent LLOs cotaining both LLSs and LLCs. Other LLOs do not contain coherences and therefore represent pure LLSs; they are obtained as linear combinations of the $\hat{B}_{0}$ operators belonging to the same global parity. First, we note that $\hat{B}_{0}(3T)$ and $\hat{B}_{0}(1T)$ correspond to the $g$--manifold, whereas $\hat{B}_{0}(2T)$ and $\hat{B}_{0}(0T)$ correspond to the $u$--manifold. In this case, the Gram--Schmidt procedure is done separately for both manifolds and yields the following operators:
\begin{align}
&\hat{L}^{(g)}_0
=
\hat{B}_0(3T)-\hat{B}_0(1T)
\nonumber\\
&=
\hat{P}(TTT)
-\frac{1}{3}
\left(
\hat{P}(TS_0S_0)+\hat{P}(S_0TS_0)+\hat{P}(S_0S_0T)
\right),
\nonumber\\
&\hat{L}^{(u)}_0
=
\hat{B}_0(0T)-\hat{B}_0(2T)
\nonumber\\
&=
\hat{P}(S_0S_0S_0)
-\frac{1}{3}
\left(
\hat{P}(S_0TT)+\hat{P}(TS_0T)+\hat{P}(TTS_0)
\right).
\label{eq:LLS-N3-B0}
\end{align}

After normalization by their corresponding norms, these operators form an orthonormal basis set of long-lived operators:
\begin{align}
&\hat{L}^{(g)}_0
=
\frac{3\sqrt{3}}{2}
\Bigl\{
\hat{P}(TTT)
\nonumber\\[-0.2em]
&\quad
-\frac{1}{3}
\bigl[
\hat{P}(TS_0S_0)
+\hat{P}(S_0TS_0)
+\hat{P}(S_0S_0T)
\bigr]
\Bigr\},
\nonumber\\[0.3em]
&\hat{L}^{(g)}_1
=
\sqrt{\frac{3}{2}}
\Bigl\{
\hat{P}(TS_0S_0)
-\hat{P}(S_0S_0T)
\Bigr\},
\nonumber\\[0.3em]
&\hat{L}^{(g)}_2
=
\frac{1}{\sqrt{2}}
\Bigl\{
\hat{P}(TS_0S_0)
-2\hat{P}(S_0TS_0)
+\hat{P}(S_0S_0T)
\Bigr\},
\nonumber\\[0.3em]
\nonumber\\[0.2em]
&\hat{L}^{(u)}_0
=
\frac{3}{2}\sqrt{\frac{3}{7}}
\Bigl\{
\hat{P}(S_0S_0S_0)
\nonumber\\[-0.2em]
&\quad
-\frac{1}{3}
\bigl[
\hat{P}(S_0TT)
+\hat{P}(TS_0T)
+\hat{P}(TTS_0)
\bigr]
\Bigr\},
\nonumber\\[0.3em]
&\hat{L}^{(u)}_1
=
\frac{3}{\sqrt{2}}
\Bigl\{
\hat{P}(S_0TT)
-\hat{P}(TTS_0)
\Bigr\},
\nonumber\\[0.3em]
&\hat{L}^{(u)}_2
=
\sqrt{\frac{3}{2}}
\Bigl\{
\hat{P}(S_0TT)
-2\hat{P}(TS_0T)
+\hat{P}(TTS_0)
\Bigr\}.
\label{eq:LLS-N3}
\end{align}
The same operators can also be expanded in the product-operator basis generated by intra-pair scalar spin products, as summarized in Table~\ref{tab:N3-LLS-product-operator-coefficients}.

The remaining operator involves imbalances between the states with different global parity: $g$ and $u$ and can be constructed according to Eq.~\eqref{eq:Lr}:
\begin{equation}
\begin{aligned}
\hat{L}_r&=\frac{1}{8\sqrt{7}}\Bigl\{
63\hat{P}(TTT)-3\hat{P}(S_0S_0S_0)
\\[-0.2em]
&+7\bigl(
\hat{P}(TS_0S_0)+\hat{P}(S_0TS_0)+\hat{P}(S_0S_0T)
\bigr)
\\[-0.2em]
&-27\bigl(
\hat{P}(S_0TT)+\hat{P}(TS_0T)+\hat{P}(TTS_0)
\bigr)
\Bigr\}.
\end{aligned}
\end{equation}

Altogether, these operators form a complete set of $2^3-1=7$ long-lived operators. For an aliphatic chain in an achiral molecule described by an $AA'MM'XX'$ spin system, the operator $\hat{L}_r$ is not experimentally accessible. Consequently, no more than $2^3-2=6$ of the operators listed in Eq.~\eqref{eq:LLS-N3} can be excited. In a chiral molecule, however, $\hat{L}_r$ can in principle become experimentally accessible.

\subsubsection{The Long-Lived States and Coherences in the Delocalized Basis}

The operators constructed above are written in the localized singlet--triplet basis. In this representation, all long-lived operators appear as population imbalances, i.e. long-lived $states$. However, for chains with more than one $-\mathrm{CH}_2-$ group, the localized singlet--triplet basis is not the eigenbasis of the $J$-coupling Hamiltonian of Eq.~\eqref{eq:H-J-aliphatic-chain}. As a result, some localized population imbalances apparently correspond to collective long-lived coherences in the delocalized basis.

\par\vspace{-1.15em}
\subsubsection*{The $\mathbf{g}$-manifold}

Here, we consider the $P(1T)$ subspace that corresponds to the $g$-manifold. As shown in Fig.~\ref{fig:2}d for $N=3$, the localized states of the type $T_0S_0S_0,S_0T_0S_0,S_0S_0T_0$  are mixed and form a separate three-dimensional subspace, therefore forming a set of delocalized states: \cite{sheberstov2025aliphatic}
\begin{align}
|\psi^{\mathrm{deloc}}_1\rangle
&=
\frac{1}{2}|T_0S_0S_0\rangle
+
\frac{1}{\sqrt{2}}|S_0T_0S_0\rangle
+
\frac{1}{2}|S_0S_0T_0\rangle,
\nonumber\\
|\psi^{\mathrm{deloc}}_2\rangle
&=
\frac{1}{\sqrt{2}}|T_0S_0S_0\rangle
-
\frac{1}{\sqrt{2}}|S_0S_0T_0\rangle,
\nonumber\\
|\psi^{\mathrm{deloc}}_3\rangle
&=
\frac{1}{2}|T_0S_0S_0\rangle
-
\frac{1}{\sqrt{2}}|S_0T_0S_0\rangle
+
\frac{1}{2}|S_0S_0T_0\rangle.
\label{eq:deloc-g-N3}
\end{align}
These wavefunctions are eigenstates for $N=3$ methylene groups in an achiral molecule, whereas all other localized states in the $g$-manifold are not mixed with each other. 

The localized populations can then be expanded as
\begin{widetext}
\begin{align}
&\hat{P}(T_0S_0S_0)
=
\frac{1}{4}\hat{P}(\psi^{\mathrm{deloc}}_1)
+
\frac{1}{2}\hat{P}(\psi^{\mathrm{deloc}}_2)
+
\frac{1}{4}\hat{P}(\psi^{\mathrm{deloc}}_3)
\nonumber\\
&
+
\frac{1}{2\sqrt{2}}
\left(
|\psi^{\mathrm{deloc}}_1\rangle\langle\psi^{\mathrm{deloc}}_2|
+
|\psi^{\mathrm{deloc}}_2\rangle\langle\psi^{\mathrm{deloc}}_1|
+
|\psi^{\mathrm{deloc}}_2\rangle\langle\psi^{\mathrm{deloc}}_3|
+
|\psi^{\mathrm{deloc}}_3\rangle\langle\psi^{\mathrm{deloc}}_2|
\right)
+
\frac{1}{4}
\left(
|\psi^{\mathrm{deloc}}_1\rangle\langle\psi^{\mathrm{deloc}}_3|
+
|\psi^{\mathrm{deloc}}_3\rangle\langle\psi^{\mathrm{deloc}}_1|
\right),
\nonumber\\[0.4em]
&\hat{P}(S_0T_0S_0)
=
\frac{1}{2}\hat{P}(\psi^{\mathrm{deloc}}_1)
+
\frac{1}{2}\hat{P}(\psi^{\mathrm{deloc}}_3)
-
\frac{1}{2}
\left(
|\psi^{\mathrm{deloc}}_1\rangle\langle\psi^{\mathrm{deloc}}_3|
+
|\psi^{\mathrm{deloc}}_3\rangle\langle\psi^{\mathrm{deloc}}_1|
\right),
\nonumber\\[0.4em]
&\hat{P}(S_0S_0T_0)
=
\frac{1}{4}\hat{P}(\psi^{\mathrm{deloc}}_1)
+
\frac{1}{2}\hat{P}(\psi^{\mathrm{deloc}}_2)
+
\frac{1}{4}\hat{P}(\psi^{\mathrm{deloc}}_3)
\nonumber\\
&
-
\frac{1}{2\sqrt{2}}
\left(
|\psi^{\mathrm{deloc}}_1\rangle\langle\psi^{\mathrm{deloc}}_2|
+
|\psi^{\mathrm{deloc}}_2\rangle\langle\psi^{\mathrm{deloc}}_1|
+
|\psi^{\mathrm{deloc}}_2\rangle\langle\psi^{\mathrm{deloc}}_3|
+
|\psi^{\mathrm{deloc}}_3\rangle\langle\psi^{\mathrm{deloc}}_2|
\right)
+
\frac{1}{4}
\left(
|\psi^{\mathrm{deloc}}_1\rangle\langle\psi^{\mathrm{deloc}}_3|
+
|\psi^{\mathrm{deloc}}_3\rangle\langle\psi^{\mathrm{deloc}}_1|
\right).
\label{eq:localized-populations-g-N3}
\end{align}

\noindent
These relations show that the long-lived operators $\hat{L}^{(g)}_1$ and $\hat{L}^{(g)}_2$
contain coherences in the delocalized basis. Explicitly,
\begin{align}
&\hat{L}^{(g)}_1
=
\sqrt{\frac{1}{6}}
\Biggl\{
\sum_{i=\pm1}
\left(
\hat{P}(T_iS_0S_0)
-
\hat{P}(S_0S_0T_i)
\right)
+
\frac{1}{\sqrt{2}}
\Bigl(
|\psi^{\mathrm{deloc}}_1\rangle
 \langle\psi^{\mathrm{deloc}}_2|
+
|\psi^{\mathrm{deloc}}_2\rangle
 \langle\psi^{\mathrm{deloc}}_1|
+
|\psi^{\mathrm{deloc}}_2\rangle
 \langle\psi^{\mathrm{deloc}}_3|
+
|\psi^{\mathrm{deloc}}_3\rangle
 \langle\psi^{\mathrm{deloc}}_2|
\Bigr)
\Biggr\},
\nonumber\\[0.4em]
&\hat{L}^{(g)}_2
=
\frac{1}{3\sqrt{2}}
\Biggl\{
\sum_{i=\pm1}
\left(
\hat{P}(T_iS_0S_0)
-
2\hat{P}(S_0T_iS_0)
+
\hat{P}(S_0S_0T_i)
\right)
+
|\psi^{\mathrm{deloc}}_1\rangle
 \langle\psi^{\mathrm{deloc}}_3|
+
|\psi^{\mathrm{deloc}}_3\rangle
 \langle\psi^{\mathrm{deloc}}_1|
\Biggr\}.
\label{eq:L-g-LLC-N3}
\end{align}
\end{widetext}

Thus, although $\hat{L}^{(g)}_1$ and $\hat{L}^{(g)}_2$ are written as population imbalances in the localized basis, they contain coherences in the delocalized eigenbasis. The observation of these long-lived coherences via polychromatic SLIC pulse sequences in achiral molecules has been reported previously. \cite{sheberstov2024collective} 

In contrast to these operators, $\hat{L}^{(g)}_0$ of Eq.~(\ref{eq:LLS-N3}) remains a pure long-lived state and does not contain coherences in the delocalized basis set since
\begin{equation}
\begin{aligned}
&\hat{P}(T_0S_0S_0)
+\hat{P}(S_0T_0S_0)
+\hat{P}(S_0S_0T_0)
\\[-0.2em]
&\quad
=
\hat{P}(\psi^{\mathrm{deloc}}_1)
+
\hat{P}(\psi^{\mathrm{deloc}}_2)
+
\hat{P}(\psi^{\mathrm{deloc}}_3).
\end{aligned}
\end{equation}

This analysis demonstrates that the distinction between LLS and LLC depends on the basis in which the long-lived operator is represented. In the localized singlet--triplet basis, all operators in Eq.~\eqref{eq:LLS-N3} appear as population imbalances. However, after transformation to the delocalized eigenbasis of the Hamiltonian, only $\hat{L}^{(g)}_0$ and $\hat{L}^{(u)}_0$ do not contain coherences, while $\hat{L}^{(g)}_{1,2}$ and $\hat{L}^{(u)}_{1,2}$ contain the contributions from both population imbalances and coherences. The consideration of the $u$-manifold can be done in the same manner and is provided in Appendix C.

\par
\vspace*{-1.0\baselineskip}
\subsection{The Long-Lived Operators for $-(\text{CH}_2)_4-$}

For $N=4$, there are five different subspaces, $P(4T), P(3T), P(2T), P(1T), P(0T)$. Two subspaces are one-dimensional:
\begin{align}
P(0T)
&=
\operatorname{span}\left\{\hat{P}(S_0S_0S_0S_0)\right\},
\nonumber\\
P(4T)
&=
\operatorname{span}\left\{\hat{P}(TTTT)\right\}.
\label{eq:P0T-P4T-N4}
\end{align}

The other subspaces are four- and six-dimensional:
\begin{align}
P(1T)
&=
\operatorname{span}
\left\{
\begin{aligned}
&
\hat{P}(TS_0S_0S_0),\quad
\hat{P}(S_0TS_0S_0),
\\[-0.2em]
&
\hat{P}(S_0S_0TS_0),\quad
\hat{P}(S_0S_0S_0T)
\end{aligned}
\right\},
\nonumber\\[0.4em]
P(2T)
&=
\operatorname{span}
\left\{
\begin{aligned}
&
\hat{P}(TTS_0S_0),\quad
\hat{P}(TS_0TS_0),
\\[-0.2em]
&
\hat{P}(TS_0S_0T),\quad
\hat{P}(S_0TTS_0),
\\[-0.2em]
&
\hat{P}(S_0TS_0T),\quad
\hat{P}(S_0S_0TT)
\end{aligned}
\right\},
\nonumber\\[0.4em]
P(3T)
&=
\operatorname{span}
\left\{
\begin{aligned}
&
\hat{P}(S_0TTT),\quad
\hat{P}(TS_0TT),
\\[-0.2em]
&
\hat{P}(TTS_0T),\quad
\hat{P}(TTTS_0)
\end{aligned}
\right\}.
\label{eq:P-manifolds-N4}
\end{align}

A very similar procedure based on the Young diagrams can be done for an arbitrary $N$ to construct the required basis of LLOs. However, its manual construction rapidly becomes complicated. For this purpose, to determine the orthogonal irreducible basis for $[N-r,r]$ diagrams we use a Python code based on symbolic calculations. Note that for $P(1T)$ and $P(3T)$ two Young diagrams are allowed: $[4,0]; [3,1]$, whereas for $P(2T)$ there are three Young diagrams: $[4,0]; [3,1]; [2,2]$.

This allows one to determine the basis set of operators that is
irreducible with respect to permutations, see Appendix D. Thus, the basis set of the long-lived operators is given by:

\begin{widetext}
\begin{equation}
\begin{aligned}
\hat{L}^{(g)}_0
&=
\frac{9}{\sqrt{82}}
\Bigl\{
\hat{B}_0(4T)-\hat{B}_0(0T)
\Bigr\},
&\qquad
\hat{L}^{(u)}_0
&=
\frac{3\sqrt{30}}{5}
\Bigl\{
\hat{B}_0(3T)-\hat{B}_0(1T)
\Bigr\},
\\
\hat{L}^{(g)}_1
&=
\frac{3\sqrt{4182}}{2788}
\Bigl\{
-81\hat{B}_0(4T)
+82\hat{B}_0(2T)
-\hat{B}_0(0T)
\Bigr\},
&
\hat{L}^{(u)}_1
&=
\frac{3\sqrt{3}}{2}\hat{B}_{1,1}(3T),
\\
\hat{L}^{(g)}_2
&=
\frac{3\sqrt{2}}{2}\hat{B}_{1,1}(2T),
&
\hat{L}^{(u)}_2
&=
\frac{3\sqrt{6}}{2}\hat{B}_{1,2}(3T),
\\
\hat{L}^{(g)}_3
&=
\frac{3}{2}\hat{B}_{1,2}(2T),
&
\hat{L}^{(u)}_3
&=
\frac{3\sqrt{6}}{2}\hat{B}_{1,3}(3T),
\\
\hat{L}^{(g)}_4
&=
\frac{3}{2}\hat{B}_{1,3}(2T),
&
\hat{L}^{(u)}_4
&=
\frac{\sqrt{3}}{2}\hat{B}_{1,1}(1T),
\\
\hat{L}^{(g)}_5
&=
\frac{3}{2}\hat{B}_{2,1}(2T),
&
\hat{L}^{(u)}_5
&=
\frac{\sqrt{6}}{2}\hat{B}_{1,2}(1T),
\\
\hat{L}^{(g)}_6
&=
\frac{\sqrt{3}}{2}\hat{B}_{2,2}(2T),
&
\hat{L}^{(u)}_6
&=
\frac{\sqrt{6}}{2}\hat{B}_{1,3}(1T).
\end{aligned}
\label{eq:LLS-N4-g-u-B}
\end{equation}
\end{widetext}
As in the cases of $N=2$ and $N=3$, the LLOs can also be expanded in the product-operator basis generated by intra-pair scalar spin products. For $N=4$, however, the explicit expressions are substantially more involved and are omitted for compactness.

The remaining long-lived operator that comprises imbalances between the states with different global intra-pair parity, see Eq.~\eqref{eq:Lr}, is given by :
\begin{align}
\hat{L}_{r}
&=
\frac{\sqrt{255}}{1360}
\Bigl\{
405\hat{B}_0(4T)
-612\hat{B}_0(3T)
\nonumber\\
&\quad
+270\hat{B}_0(2T)
-68\hat{B}_0(1T)
+5\hat{B}_0(0T)
\Bigr\}.
\label{eq:LLS-N4-remaining-B}
\end{align}

We emphasize that, for an aliphatic chain in an achiral molecule, where the two protons within each $-\mathrm{CH}_2-$ group are chemically equivalent, the conserved global intra-pair parity prevents experimental excitation of $\hat{L}_{r}$. In a chiral molecule, the two methylene protons may become chemically non-equivalent, so that this parity is no longer conserved and $\hat{L}_{r}$ can, in principle, become experimentally accessible. 

Thus, for $N=4$, the complete long-lived subspace contains $2^4-1=15$ non-trivial operators. In an achiral molecule, however, at most $2^4-2=14$ of these operators can be excited; the complete set of experimentally accessible operators is given in Eq.~\eqref{eq:LLS-N4-g-u-B}. These experimentally accessible operators have a well-defined global $g$ or $u$ parity, whereas $\hat{L}_{r}$ involves a population imbalance between the $g$ and $u$ manifolds. In a chiral molecule, where global intra-pair parity is not conserved, this restriction is lifted and all 15 non-trivial long-lived operators can, in principle, be accessed.

It is important to emphasize that some of the operators written above as population imbalances contain coherences when expressed in the delocalized basis, as for the previously considered case $N=~2,3$. For $N=4$, the operators $\hat{L}^{(u)}_1,\ldots,\hat{L}^{(u)}_6$ originate from the subspaces with one or three triplet labels and therefore involve delocalized combinations of states such as
\begin{equation}
\nonumber
T_0S_0S_0S_0, S_0T_0S_0S_0, S_0S_0T_0S_0, S_0S_0S_0T_0,
\end{equation}
or their three-triplet counterparts.

Similarly, the operators $\hat{L}^{(g)}_2,\ldots,\hat{L}^{(g)}_6$ originate from the two-triplet subspace and involve delocalized combinations of states such as
\begin{equation}
\begin{aligned}
\nonumber
&
T_0T_0S_0S_0,\quad
T_0S_0T_0S_0,\quad
T_0S_0S_0T_0,
\nonumber\\[-0.2em]
&
S_0T_0T_0S_0,\quad
S_0T_0S_0T_0,\quad
S_0S_0T_0T_0.
\end{aligned}
\end{equation}
Thus, these operators should be understood not only as long-lived states in the localized basis, but also as long-lived coherences in the delocalized eigenbasis of the Hamiltonian. In contrast, $\hat{L}^{(g)}_0$, $\hat{L}^{(g)}_1$, and $\hat{L}^{(u)}_0$ remain purely LLSs, since they are constructed from invariant combinations that do not generate coherences upon transformation to the delocalized eigenbasis (for achiral molecules considered in this work).

An identical procedure can be done for a chain with an arbitrary number $N$ of $-\text{CH}_2$ groups. 

\section{Methods}

All analytical expressions for the long-lived operators for $N=4$ were generated using an in-house Python code. Our implementation is not restricted to $N=4$ and can be used to construct the corresponding symbolic LLO basis for an arbitrary number of $-\mathrm{CH}_2-$ groups, limited only by the rapidly increasing size of the operator space.

The code symbolically constructs the local singlet--triplet population basis for each $-\mathrm{CH}_2-$ group and then forms the product-operator basis for a chain containing $N$ groups. The resulting operators are sorted into triplet-number manifolds $P(MT)$ and decomposed into permutation-adapted components using Young symmetrizers. Within each irreducible component, linearly independent operators are selected and orthogonalized by the Gram--Schmidt procedure.

The fully symmetric components $\hat B_0(MT)$ are subsequently combined within the corresponding global intra-pair parity manifolds to obtain traceless LLS operators that do not contain coherences. The non-symmetric components $\hat B_{r,\alpha}(MT)$ with $r\neq0$ are already traceless and are retained as long-lived operators associated, in the delocalized Hamiltonian eigenbasis, with long-lived coherences. All operators are finally normalized to unit norm.

\section{Conclusions}

In this work, we have shown that delocalized long-lived states and collective long-lived coherences in aliphatic chains arise naturally from Redfield relaxation theory. Starting from the intra-pair dipole--dipole relaxation superoperator, we derived the local zero-eigenvalue subspace of a single $-\mathrm{CH}_2-$ group and showed that it is spanned by the singlet population and the equally weighted triplet population, namely 
$\hat{P}\!\left(S_0\right)
=
\left|
S_0
\right\rangle
\left\langle
S_0
\right|$ and $\hat{P}\!\left(T\right)
=~
\frac{1}{3}
\sum_{i=-1}^{1}
|
T_{i}
\rangle
\langle
T_{i}
|$. These local operators provide the elementary building block for the construction of long-lived operators in aliphatic chains containing an arbitrary number $N$ of $-\mathrm{CH}_2-$ groups.

We then developed a general construction of the long-lived operator basis, which is given by the product of a certain number of $\hat{P}(S_0)$ and $\hat{P}(T)$ projectors. These products were classified into triplet-number manifolds $P(MT)$ and decomposed into irreducible representations of the permutation group. This construction yields an orthonormal set of long-lived operators. We show that for achiral molecules, the fully permutation-symmetric basis components give purely long-lived population imbalances, i.e. long-lived states, whereas the non-trivial permutation components give operators that appear as population imbalances in the localized singlet--triplet basis but also contain long-lived coherences in the delocalized eigenbasis of the coherent Hamiltonian.

The resulting counting shows that the full zero-eigenvalue subspace contains $2^N-1$ non-trivial long-lived operators. In achiral molecules, conservation of the global intra-pair parity restricts the number of experimentally excitable operators to at most $2^N-2$. The remaining operator involves a population imbalance between states of different global intra-pair parity and is therefore inaccessible. In chiral molecules, this restriction is lifted, allowing all $2^N-1$ non-trivial operators to become experimentally accessible in principle.

Finally, the present analysis connects long-lived spin order with collective spin dynamics in aliphatic chains: it provides a natural basis for describing delocalized long-lived operators, thereby linking the theory of LLS and LLC to spin-wave-like excitations in multi-spin systems. In particular, for chains containing many $-\mathrm{CH}_2-$ groups, the present operator basis should make it possible to identify which long-lived operators are populated most efficiently by polychromatic SLIC-type pulse sequences. A detailed analysis of SLIC excitation pathways in chains of different length, together with the connection between these long-lived operators and spin-wave dynamics, will be presented in forthcoming work.

\begin{acknowledgments}
The authors acknowledge l’Agence Nationale de la Recherche (ANR) on the project THROUGH-NMR (grant no. ANR-24-CE93-0011-01).
\end{acknowledgments}

\section*{Author Declarations}

\subsection*{Conflict of Interest}

The authors have no conflicts to disclose.

\subsection*{Author Contributions}

Danil A. Markelov: Conceptualization, Investigation, Methodology, Formal analysis, Writing -- original draft, Software.
Kirill F. Sheberstov: Conceptualization, Investigation, Methodology, Formal analysis, Writing -- original draft, Supervision, Project Administration.

\section*{Data Availability}

The Python code constructing the long-lived operators in the aliphatic chain of an arbitrary length is available at https://doi.org/10.5281/zenodo.21835882. No external datasets were generated or analyzed during the current study.

\appendix

\section{Glossary}
\label{app:glossary}

\begin{description}

\item[\textbf{Long-lived operator (LLO)}]
An operator that belongs to the zero-eigenvalue subspace of the dominant
intra-pair dipole--dipole relaxation superoperator, $\hat{\hat{\mathbf{R}}}_{\text{intra}}$.

\item[\textbf{Delocalized state}]
A state formed as a linear combination of localized singlet--triplet product
states involving different \(-\mathrm{CH}_2-\) groups. For example,
\(\left|S_0T_0\right\rangle\) is a localized separable state of two
\(-\mathrm{CH}_2-\) groups, since it can be written as a product of a state
localized on the first group and a state localized on the second group,
$
\left|S_0T_0\right\rangle
=
\left|S_0\right\rangle_1
\left|T_0\right\rangle_2 .
$
In contrast, for example,
\[
\frac{\left|S_0T_0\right\rangle
+
\left|T_0S_0\right\rangle}{\sqrt{2}}
\]
is a delocalized state. It cannot be factorized into a product
\(\left|\chi\right\rangle_1\left|\eta\right\rangle_2\) of states localized on
the individual \(-\mathrm{CH}_2-\) groups.

\item[\textbf{Long-lived state (LLS) component}]
The diagonal part of a long-lived operator when it is written
in the delocalized eigenbasis. If a long-lived operator is purely diagonal in
this basis, it represents a pure LLS.

\item[\textbf{Long-lived coherence (LLC) component}]
The off-diagonal part of a long-lived operator when it is
written in the delocalized eigenbasis. In aliphatic chains, LLCs do
not appear as isolated operators. Rather, they occur as a part of
a zero-eigenvalue mode that also contains populations. Thus, the
term LLC refers to the coherence component of such a long-lived operator.

\end{description}

\section{The Irreducible Basis Set for $-(\text{CH}_2)_3-$}

For $N=3$, there are four relevant subspaces:
$P(0T)$, $P(1T)$, $P(2T)$, and $P(3T)$. The subspaces $P(0T)$ and $P(3T)$ are one-dimensional:
\begin{equation}
\begin{aligned}
P(0T)
&=
\operatorname{span}\left\{\hat{P}(S_0S_0S_0)\right\},\\
P(3T)
&=
\operatorname{span}\left\{\hat{P}(TTT)\right\}.
\end{aligned}
\end{equation}

The subspaces $P(1T)$ and $P(2T)$ are three-dimensional:
\begin{equation}
\begin{aligned}
P(1T)
&=
\operatorname{span}
\left\{
\hat{P}(TS_0S_0),\hat{P}(S_0TS_0),\hat{P}(S_0S_0T)
\right\},\\
P(2T)
&=
\operatorname{span}
\left\{
\hat{P}(S_0TT),\hat{P}(TS_0T),\hat{P}(TTS_0)
\right\}.
\end{aligned}
\end{equation}

Let us first consider the subspace $P(1T)$. The treatment of $P(2T)$ is completely analogous. The representation carried by $P(1T)$ decomposes into two irreducible Young components, $[3,0]$ and $[2,1]$.

\subsubsection{The Young diagram $\mathbf{[3,0]}$ for $\mathbf{P(1T)}$}

The basis operator symmetric under all permutations, r = 0, is obtained from the only allowed tableau
\begin{equation}
t=
\begin{array}{ccc}
1 & 2 & 3
\end{array}.
\end{equation}

This tableau gives the following Young superoperator:
\begin{equation}
\hat{\hat{Y}}_t
=
\hat{\hat{E}}
+
\hat{\hat{\Pi}}_{12}
+
\hat{\hat{\Pi}}_{13}
+
\hat{\hat{\Pi}}_{23}
+
\hat{\hat{\Pi}}_{12}\hat{\hat{\Pi}}_{23}
+
\hat{\hat{\Pi}}_{12}\hat{\hat{\Pi}}_{13}.
\end{equation}

Applying $\hat{\hat{Y}}_t$ to, for example, $\hat{P}(TS_0S_0)$ gives
\begin{equation}
2\left(
\hat{P}(TS_0S_0)+\hat{P}(S_0TS_0)+\hat{P}(S_0S_0T)
\right).
\end{equation}

Thus, after normalization so that the trace is equal to unity, we obtain a single permutation-symmetric component associated with the $[3,0]$ diagram:
\begin{equation}
\hat{B}_0(1T)
=
\frac{1}{3}
\left(
\hat{P}(TS_0S_0)+\hat{P}(S_0TS_0)+\hat{P}(S_0S_0T)
\right).
\end{equation}

\subsubsection{The Young diagram $\mathbf{[2,1]}$ for $\mathbf{P(1T)}$}

For the diagram $[2,1]$, two standard tableaux are allowed, and the corresponding irreducible representation is two-dimensional. Let us first consider the tableau
\begin{equation}
t=
\begin{array}{cc}
1 & 2\\
3 &
\end{array}.
\end{equation}

For this tableau, the row symmetrization superoperator is
\begin{equation}
\hat{\hat{S}}_t
=
\hat{\hat{E}}
+
\hat{\hat{\Pi}}_{12},
\end{equation}
whereas the column antisymmetrization superoperator is
\begin{equation}
\hat{\hat{A}}_t
=
\hat{\hat{E}}
-
\hat{\hat{\Pi}}_{13}.
\end{equation}

\begin{table*}[t]
\caption{
Expansion coefficients of the long-lived operators in the $g/u$--manifolds for $N=3$ in the scalar-product-operator basis set
$\hat{X}^{(1,2)}=\hat{\mathbf I}^{(1)}\cdot\hat{\mathbf I}^{(2)}$,
$\hat{X}^{(3,4)}=\hat{\mathbf I}^{(3)}\cdot\hat{\mathbf I}^{(4)}$, and
$\hat{X}^{(5,6)}=\hat{\mathbf I}^{(5)}\cdot\hat{\mathbf I}^{(6)}$. The index pairs $(1,2)$; $(3,4)$, and $(5,6)$ correspond to the first, second, and third $-\mathrm{CH}_2-$ groups, respectively. All operators are normalized to unit Hilbert--Schmidt norm, i.e., $||\hat{L}||=1$.
}
\centering
\scriptsize
\renewcommand{\arraystretch}{3}
\setlength{\tabcolsep}{5pt}
\begin{tabular}{c c c c c c c c}
\hline
Operator
&
$\hat{X}^{(1,2)}$
&
$\hat{X}^{(3,4)}$
&
$\hat{X}^{(5,6)}$
&
$\hat{X}^{(1,2)}\hat{X}^{(3,4)}$
&
$\hat{X}^{(1,2)}\hat{X}^{(5,6)}$
&
$\hat{X}^{(3,4)}\hat{X}^{(5,6)}$
&
$\hat{X}^{(1,2)}\hat{X}^{(3,4)}\hat{X}^{(5,6)}$
\\
\hline

$\hat L^{(g)}_0$
&
$\dfrac{\sqrt{3}}{12}$
&
$\dfrac{\sqrt{3}}{12}$
&
$\dfrac{\sqrt{3}}{12}$
&
$0$
&
$0$
&
$0$
&
$-\dfrac{4\sqrt{3}}{9}$
\\

$\hat L^{(g)}_1$
&
$\dfrac{\sqrt{6}}{24}$
&
$0$
&
$-\dfrac{\sqrt{6}}{24}$
&
$-\dfrac{\sqrt{6}}{6}$
&
$0$
&
$\dfrac{\sqrt{6}}{6}$
&
$0$
\\

$\hat L^{(g)}_2$
&
$\dfrac{\sqrt{2}}{24}$
&
$-\dfrac{\sqrt{2}}{12}$
&
$\dfrac{\sqrt{2}}{24}$
&
$\dfrac{\sqrt{2}}{6}$
&
$-\dfrac{\sqrt{2}}{3}$
&
$\dfrac{\sqrt{2}}{6}$
&
$0$
\\

$\hat L^{(u)}_0$
&
$-\dfrac{\sqrt{21}}{84}$
&
$-\dfrac{\sqrt{21}}{84}$
&
$-\dfrac{\sqrt{21}}{84}$
&
$\dfrac{4\sqrt{21}}{63}$
&
$\dfrac{4\sqrt{21}}{63}$
&
$\dfrac{4\sqrt{21}}{63}$
&
$-\dfrac{4\sqrt{21}}{21}$
\\

$\hat L^{(u)}_1$
&
$-\dfrac{\sqrt{2}}{8}$
&
$0$
&
$\dfrac{\sqrt{2}}{8}$
&
$-\dfrac{\sqrt{2}}{6}$
&
$0$
&
$\dfrac{\sqrt{2}}{6}$
&
$0$
\\

$\hat L^{(u)}_2$
&
$-\dfrac{\sqrt{6}}{24}$
&
$\dfrac{\sqrt{6}}{12}$
&
$-\dfrac{\sqrt{6}}{24}$
&
$\dfrac{\sqrt{6}}{18}$
&
$-\dfrac{\sqrt{6}}{9}$
&
$\dfrac{\sqrt{6}}{18}$
&
$0$
\\

\hline
\end{tabular}
\label{tab:N3-LLS-product-operator-coefficients}
\end{table*}

Using the convention $\hat{\hat{Y}}_t=\hat{\hat{A}}_t\hat{\hat{S}}_t$, the total Young superoperator is
\begin{equation}
\hat{\hat{Y}}_t
=
\hat{\hat{A}}_t\hat{\hat{S}}_t
=
\hat{\hat{E}}
+
\hat{\hat{\Pi}}_{12}
-
\hat{\hat{\Pi}}_{13}
-
\hat{\hat{\Pi}}_{13}\hat{\hat{\Pi}}_{12}.
\end{equation}

Application of $\hat{\hat{Y}}_t$ to $\hat{P}(TS_0S_0)$ gives
\begin{equation}
\hat{O}_1
=
\hat{\hat{Y}}_t\hat{P}(TS_0S_0)
=
\hat{P}(TS_0S_0)-\hat{P}(S_0S_0T).
\end{equation}
This operator can be used as one of the basis operators for the LLS construction.

Applying the same superoperator to $\hat{P}(S_0TS_0)$ gives
\begin{equation}
\hat{\hat{Y}}_t\hat{P}(S_0TS_0)
=
\hat{P}(TS_0S_0)-\hat{P}(S_0S_0T),
\end{equation}
which is linearly dependent on $\hat{O}_1$.

We therefore consider the second standard tableau:
\begin{equation}
t=
\begin{array}{cc}
1 & 3\\
2 &
\end{array}.
\end{equation}

For this tableau, the Young superoperator is
\begin{equation}
\hat{\hat{Y}}_t
=
\hat{\hat{A}}_t\hat{\hat{S}}_t
=
\hat{\hat{E}}
+
\hat{\hat{\Pi}}_{13}
-
\hat{\hat{\Pi}}_{12}
-
\hat{\hat{\Pi}}_{12}\hat{\hat{\Pi}}_{13}.
\end{equation}

Application to $\hat{P}(TS_0S_0)$ gives
\begin{equation}
\hat{O}_2
=
\hat{\hat{Y}}_t\hat{P}(TS_0S_0)
=
\hat{P}(TS_0S_0)-\hat{P}(S_0TS_0).
\end{equation}

The operators $\hat{O}_1$ and $\hat{O}_2$ are linearly independent, but they are not orthogonal. Applying the Gram--Schmidt procedure to these two operators gives the following orthogonal basis for the $[2,1]$ irreducible component:
\begin{equation}
\begin{aligned}
\hat{B}_{1,1}(1T)
&=
\hat{P}(TS_0S_0)-\hat{P}(S_0S_0T),\\
\hat{B}_{1,2}(1T)
&=
\hat{P}(TS_0S_0)-2\hat{P}(S_0TS_0)+\hat{P}(S_0S_0T).
\end{aligned}
\end{equation}

Similarly, in the subspace $P(2T)$, one obtains three additional basis operators:
\begin{equation}
\begin{aligned}
&\hat{B}_0(2T)
=
\frac{1}{3}
\left(
\hat{P}(S_0TT)+\hat{P}(TS_0T)+\hat{P}(TTS_0)
\right),\\
&\hat{B}_{1,1}(2T)
=
\hat{P}(S_0TT)-\hat{P}(TTS_0),\\
&\hat{B}_{1,2}(2T)
=
\hat{P}(S_0TT)-2\hat{P}(TS_0T)+\hat{P}(TTS_0).
\end{aligned}
\end{equation}

\section{Long-Lived Operators in the Delocalized $\mathbf{u}$-Manifold Basis}

The same argument applies to the antisymmetric, or the $u$--manifold. In this case, we consider the $P(2T)$ subspace. As shown in Fig.~\ref{fig:2}c, there are five different sub-blocks in this subspace: four of them are $2\times2$, whereas one of them is $3\times3$. 

The $3\times3$ sub-block is spanned by $S_0T_0T_0$, $T_0S_0T_0$, and $T_0T_0S_0$. The mixing between them results in the following delocalized states: \cite{sheberstov2025aliphatic}
\begin{align}
|\phi^{\mathrm{deloc}}_1\rangle
&=
\frac{1}{2}|S_0T_0T_0\rangle
+
\frac{1}{\sqrt{2}}|T_0S_0T_0\rangle
+
\frac{1}{2}|T_0T_0S_0\rangle,
\nonumber\\
|\phi^{\mathrm{deloc}}_2\rangle
&=
\frac{1}{\sqrt{2}}|S_0T_0T_0\rangle
-
\frac{1}{\sqrt{2}}|T_0T_0S_0\rangle,
\nonumber\\
|\phi^{\mathrm{deloc}}_3\rangle
&=
\frac{1}{2}|S_0T_0T_0\rangle
-
\frac{1}{\sqrt{2}}|T_0S_0T_0\rangle
+
\frac{1}{2}|T_0T_0S_0\rangle.
\label{eq:deloc-u-N3}
\end{align}

Consideration of four two-dimensional subspaces leads to the following delocalized eigenfunctions
\begin{align}
|\phi^{\mathrm{deloc}}_{a_{\pm1}}\rangle
&=
\frac{|T_{\pm1}T_0S_0\rangle+|T_{\pm1}S_0T_0\rangle}{\sqrt{2}},
\nonumber\\
|\phi^{\mathrm{deloc}}_{b_{\pm1}}\rangle
&=
\frac{|T_{\pm1}T_0S_0\rangle-|T_{\pm1}S_0T_0\rangle}{\sqrt{2}},
\label{eq:deloc-u-N3-ab}
\end{align}
and
\begin{align}
|\phi^{\mathrm{deloc}}_{c_{\pm1}}\rangle
&=
\frac{|T_0S_0T_{\pm1}\rangle+|S_0T_0T_{\pm1}\rangle}{\sqrt{2}},
\nonumber\\
|\phi^{\mathrm{deloc}}_{d_{\pm1}}\rangle
&=
\frac{|T_0S_0T_{\pm1}\rangle-|S_0T_0T_{\pm1}\rangle}{\sqrt{2}}.
\label{eq:deloc-u-N3-cd}
\end{align}
For compactness, we also define the corresponding coherences
\begin{align}
\hat C^{(ab)}_{i}
&=
|\phi^{\mathrm{deloc}}_{a_i}\rangle
\langle\phi^{\mathrm{deloc}}_{b_i}|
+
|\phi^{\mathrm{deloc}}_{b_i}\rangle
\langle\phi^{\mathrm{deloc}}_{a_i}|,
\nonumber\\
\hat C^{(cd)}_{i}
&=
|\phi^{\mathrm{deloc}}_{c_i}\rangle
\langle\phi^{\mathrm{deloc}}_{d_i}|
+
|\phi^{\mathrm{deloc}}_{d_i}\rangle
\langle\phi^{\mathrm{deloc}}_{c_i}|,
\quad i=\pm1 .
\end{align}

In complete analogy with the $P(1T)$ subspace, the localized
population operators
\begin{equation}
\begin{aligned}
\nonumber
&
\hat{P}(S_0T_0T_0),\quad
\hat{P}(T_0S_0T_0),\quad
\hat{P}(T_0T_0S_0),
\\[-0.2em]
&
\hat{P}(T_{\pm1}T_0S_0),\quad
\hat{P}(T_{\pm1}S_0T_0),
\\[-0.2em]
&
\hat{P}(T_0S_0T_{\pm1}),\quad
\hat{P}(S_0T_0T_{\pm1})
\end{aligned}
\end{equation}
contain both populations and coherences in the delocalized basis.
For $\hat{P}(S_0T_0T_0)$, $\hat{P}(T_0S_0T_0)$, and
$\hat{P}(T_0T_0S_0)$, one can use
Eq.~(\ref{eq:localized-populations-g-N3}) directly. Similar
transformation properties can be obtained for the other delocalized
states,
$|\phi^{\mathrm{deloc}}_{a_{\pm1}}\rangle$--%
$|\phi^{\mathrm{deloc}}_{d_{\pm1}}\rangle$.

In the delocalized basis set, the normalized operators $\hat{L}^{(u)}_1$ and $\hat{L}^{(u)}_2$ can be written as
\begin{flalign}
&\hat{L}^{(u)}_1
=
\frac{1}{3\sqrt{2}}
\Biggl\{
\sum_{i,j=\pm1}
\left(
\hat{P}(S_0T_iT_j)
-
\hat{P}(T_iT_jS_0)
\right)
&&\nonumber\\
&
+
\sum_{i=\pm1}
\left(
\hat{P}(S_0T_iT_0)
-
\frac{
\hat{P}(\phi^{\mathrm{deloc}}_{a_i})
+
\hat{P}(\phi^{\mathrm{deloc}}_{b_i})
}{2}
-
\frac{1}{2}\hat{C}^{(ab)}_i
\right)
&&\nonumber\\
&
+
\sum_{i=\pm1}
\left(
\frac{
\hat{P}(\phi^{\mathrm{deloc}}_{c_i})
+
\hat{P}(\phi^{\mathrm{deloc}}_{d_i})
}{2}
-
\frac{1}{2}\hat{C}^{(cd)}_i
-
\hat{P}(T_0T_iS_0)
\right)
&&\nonumber\\
&
+
\frac{1}{\sqrt{2}}
\Bigl(
|\phi^{\mathrm{deloc}}_1\rangle
 \langle\phi^{\mathrm{deloc}}_2|
+
|\phi^{\mathrm{deloc}}_2\rangle
 \langle\phi^{\mathrm{deloc}}_1|
&&\nonumber\\
&\qquad
+
|\phi^{\mathrm{deloc}}_2\rangle
 \langle\phi^{\mathrm{deloc}}_3|
+
|\phi^{\mathrm{deloc}}_3\rangle
 \langle\phi^{\mathrm{deloc}}_2|
\Bigr)
\Biggr\},
&&\nonumber\\[0.4em]
&\hat{L}^{(u)}_2 
=
&&\nonumber\\
&
\frac{1}{3\sqrt{6}}
\Biggl\{
\sum_{i,j=\pm1}
\left(
\hat{P}(S_0T_iT_j)
-
2\hat{P}(T_iS_0T_j)
+
\hat{P}(T_iT_jS_0)
\right)
&&\nonumber\\
&
+
\sum_{i=\pm1}
\left(
\hat{P}(S_0T_iT_0)
-
\frac{
\hat{P}(\phi^{\mathrm{deloc}}_{a_i})
+
\hat{P}(\phi^{\mathrm{deloc}}_{b_i})
}{2}
+
\frac{3}{2}\hat{C}^{(ab)}_i
\right)
&&\nonumber\\
&
+
\sum_{i=\pm1}
\left(
\hat{P}(T_0T_iS_0)
-
\frac{
\hat{P}(\phi^{\mathrm{deloc}}_{c_i})
+
\hat{P}(\phi^{\mathrm{deloc}}_{d_i})
}{2}
-
\frac{3}{2}\hat{C}^{(cd)}_i
\right)
&&\nonumber\\
&
+
|\phi^{\mathrm{deloc}}_1\rangle
 \langle\phi^{\mathrm{deloc}}_3|
+
|\phi^{\mathrm{deloc}}_3\rangle
 \langle\phi^{\mathrm{deloc}}_1|
\Biggr\}.
&&\label{eq:L-u-LLC-N3}
\end{flalign}

Therefore, it is clear that $\hat{L}^{(u)}_1$ and $\hat{L}^{(u)}_2$ contain populations and coherences in the delocalized eigenbasis, whereas $\hat{L}^{(u)}_0$ of Eq.~(\ref{eq:LLS-N3}) remains a pure long-lived population imbalance since it is a permutation-invariant operator.

\section{The Irreducible Basis Set for $-(\text{CH}_2)_4-$}

The irreducible basis set for $N=4$ is given by:
\begin{widetext}
\begin{align}
P(0T):\qquad
& \hat{B}_0(0T)=\hat{P}(S_0S_0S_0S_0),
\nonumber\\[1mm]
P(1T):\qquad
& \hat{B}_0(1T)=\frac{1}{4}
\Bigl(
\hat{P}(TS_0S_0S_0)
+\hat{P}(S_0TS_0S_0)
+\hat{P}(S_0S_0TS_0)
+\hat{P}(S_0S_0S_0T)
\Bigr),
\nonumber\\
& \hat{B}_{1,1}(1T)
=
\hat{P}(TS_0S_0S_0)
-\hat{P}(S_0TS_0S_0)
-\hat{P}(S_0S_0TS_0)
+\hat{P}(S_0S_0S_0T),
\nonumber\\
& \hat{B}_{1,2}(1T)
=
\hat{P}(S_0TS_0S_0)
-\hat{P}(S_0S_0TS_0),
\nonumber\\
& \hat{B}_{1,3}(1T)
=
\hat{P}(TS_0S_0S_0)
-\hat{P}(S_0S_0S_0T),
\nonumber\\[1mm]
P(2T):\qquad
& \hat{B}_0(2T)=\frac{1}{6}
\Bigl(
\hat{P}(TTS_0S_0)
+\hat{P}(TS_0TS_0)
+\hat{P}(TS_0S_0T)
+\hat{P}(S_0TTS_0)
+\hat{P}(S_0TS_0T)
+\hat{P}(S_0S_0TT)
\Bigr),
\nonumber\\
& \hat{B}_{1,1}(2T)
=
\hat{P}(TS_0S_0T)
-\hat{P}(S_0TTS_0),
\nonumber\\
& \hat{B}_{1,2}(2T)
=
\hat{P}(TTS_0S_0)
-\hat{P}(TS_0TS_0)
+\hat{P}(S_0TS_0T)
-\hat{P}(S_0S_0TT),
\nonumber\\
& \hat{B}_{1,3}(2T)
=
\hat{P}(TTS_0S_0)
+\hat{P}(TS_0TS_0)
-\hat{P}(S_0TS_0T)
-\hat{P}(S_0S_0TT),
\nonumber\\
& \hat{B}_{2,1}(2T)
=
\hat{P}(TS_0TS_0)
-\hat{P}(TS_0S_0T)
-\hat{P}(S_0TTS_0)
+\hat{P}(S_0TS_0T),
\nonumber\\
& \hat{B}_{2,2}(2T)
=
2\hat{P}(TTS_0S_0)
-\hat{P}(TS_0TS_0)
-\hat{P}(TS_0S_0T)
-\hat{P}(S_0TTS_0)
-\hat{P}(S_0TS_0T)
+2\hat{P}(S_0S_0TT),
\nonumber\\[1mm]
P(3T):\qquad
& \hat{B}_0(3T)=\frac{1}{4}
\Bigl(
\hat{P}(TTTS_0)
+\hat{P}(TTS_0T)
+\hat{P}(TS_0TT)
+\hat{P}(S_0TTT)
\Bigr),
\nonumber\\
& \hat{B}_{1,1}(3T)
=
\hat{P}(TTTS_0)
-\hat{P}(TTS_0T)
-\hat{P}(TS_0TT)
+\hat{P}(S_0TTT),
\nonumber\\
& \hat{B}_{1,2}(3T)
=
\hat{P}(TTS_0T)
-\hat{P}(TS_0TT),
\nonumber\\
& \hat{B}_{1,3}(3T)
=
\hat{P}(TTTS_0)
-\hat{P}(S_0TTT),
\nonumber\\[1mm]
P(4T):\qquad
& \hat{B}_0(4T)=\hat{P}(TTTT).
\label{eq:B-basis-N4}
\end{align}
\end{widetext}

All operators except for $\hat{B}_0(MT)$ already represent the required
traceless basis set. To construct the basis set from $\hat{B}_0$, we
consider the $g$-- and $u$-- manifolds separately. The $g$--manifold contains
$\hat{B}_0(4T)$, $\hat{B}_0(2T)$, and $\hat{B}_0(0T)$, whereas the
$u$--manifold contains $\hat{B}_0(3T)$ and $\hat{B}_0(1T)$. Application
of the Gram--Schmidt procedure to each set, with the requirement
$\mathrm{Tr}\{\hat{L}\}=0$, produces the complete orthonormal basis set, see Eq.~\eqref{eq:LLS-N4-g-u-B}.

\clearpage
\section*{References}
\bibliography{references}

\end{document}